\renewcommand\L{\Lambda}

\newcommand{\diracslash}[1]{#1\llap{/\kern2pt}}

\newcommand{\be}{\begin{equation}}
\newcommand{\ee}{\end{equation}}
\newcommand{\bea}{\begin{eqnarray}}
\newcommand{\eea}{\end{eqnarray}}
\newcommand{\ba}[1]{\begin{array}{#1}}
\newcommand{\ea}{\end{array}}

\newcommand{\bt}{\begin{tabular}}
\newcommand{\et}{\end{tabular}}

\newcommand{\beas}{\begin{eqnarray*}}
\newcommand{\eeas}{\end{eqnarray*}}

\documentclass[preprint,prd,aps,floats,nofootinbib,floatfix]{revtex4}

\DeclareSymbolFont{rsfs}{U}{rsfs}{m}{n}
\DeclareSymbolFontAlphabet{\mathrsfs}{rsfs}
\usepackage{graphicx}
\usepackage{multirow}

\usepackage{aliascnt}

\usepackage{pgf}
\usepackage{csquotes}
\usepackage{pdfpages}
\usepackage{amsmath}
\usepackage[amsmath,thmmarks]{ntheorem}

\usepackage{graphicx}
\usepackage{cleveref}

\begin{document}

%\title{Pseudoscalar D and B mesons in hot and dense asymmetric hadronic medium}

\title{Masses and decay widths of  scalar $D_0$ and $D_{s0}$ mesons in strange hadronic medium}
%%%%%%%%%%%%%%%%%%%Ds%%%%%%%%%%%%%%%%%%%%%%%%%%%%%%%%%%%%%%%%%%%%%%

\author{Rahul Chhabra}
\email{rahulchhabra@ymail.com,}
\affiliation{Department of Physics, Dr. B R Ambedkar National Institute of Technology Jalandhar,
 Jalandhar -- 144011,Punjab, India}

\author{Arvind Kumar}
\email{iitd.arvind@gmail.com, kumara@nitj.ac.in}
\affiliation{Department of Physics, Dr. B R Ambedkar National Institute of Technology Jalandhar,
 Jalandhar -- 144011,Punjab, India}

\def\be{\begin{equation}}
\def\ee{\end{equation}}
\def\bearr{\begin{eqnarray}}
\def\eearr{\end{eqnarray}}
\def\zbf#1{{\bf {#1}}}
\def\bfm#1{\mbox{\boldmath $#1$}}
\def\hf{\frac{1}{2}}
\def\kp{\zbf k+\frac{\zbf q}{2}}
\def\km{-\zbf k+\frac{\zbf q}{2}}
\def\hwo{\hat\omega_1}
\def\hwt{\hat\omega_2}

\begin{abstract}

Masses and decay constants of scalar $D_0$ and $D_{s0}$ mesons in  isospin asymmetric strange hadronic matter at finite temperature are evaluated using QCD sum rules and chiral SU(3) model. In-medium light quark condensates, $\left\langle \bar{u}u\right\rangle_{\rho_{B}}$ and $\left\langle \bar{d}d\right\rangle_{\rho_{B}}$, the strange quark condensates, $\left\langle \bar{s}s\right\rangle_{\rho_{B}}$, and the gluon condensates, $\left\langle  \frac{\alpha_{s}}{\pi} {G^a}_{\mu\nu} {G^a}^{\mu\nu}
\right\rangle_{\rho_{B}}$, needed in QCD sum rule calculations are evaluated using chiral SU(3) model.  As an application, we calculate the in-medium partial decay width of scalar $D_0$ ($D_{s0}$) meson decaying to $D$ + $\pi$ ($D_s$+$\pi$) pseudoscalar mesons using $^3 P_0$ model.  The medium effects in their decay widths are assimilated through the modification in the masses of these mesons. These results may be helpful to understand the possible outcomes of the future experiments like CBM and PANDA under the FAIR facility
where the study of charmed hadrons is one of major goal. 

\textbf{Keywords:} Dense hadronic matter, strangeness fraction,
 heavy-ion collisions, effective chiral model, QCD sum rules,  heavy mesons.

PACS numbers : -14.40.Lb ,-14.40.Nd,13.75.Lb
\end{abstract}

\maketitle
\section{Introduction}
In-medium study of $D$ meson has been an object of intense study \cite{tsu,tolo1,hoff,CE,CE1,tolo,haya2000,hilger2009,arvinds2,pathaks,tolos,wang2,arvinds}, because of the observation of its enhanced yield \cite{open_charm_en1,open_charm_en2,open_charm_en3} and also its possible consequence on the  yield of higher charmonium states observed in heavy-ion collision (HIC) experiments \cite{NA38,NA50pb,NA60,rhic}.  Firstly, it was proposed by Matsui and Satz that the decrease in the yield of $J/\psi$ state in HICs due to color screening effect should be considered as a probe of the production of the state existed in early universe, i.e., Quark Gluon Plasma (QGP) \cite{mats}. 
 Since then, imperative  results in the favour of $J/\psi$ suppression were observed at CERN SPS
% experiment NA38 \cite{NA38}, NA50 \cite{NA50pb},  NA60 \cite{NA60}
 and in the RHIC experiment \cite{NA38,NA50pb,NA60,rhic}.
The statistical recombination of primordially produced charm 
quark pairs may lead to the increase in the yield of $J/\psi$ mesons and this picture is more important at LHC energies \cite{LHC1,LHC2}.
The behaviour of in-medium masses and spectral width of $D$ mesons will play an important role on the final yield of charmonium. 
If the drop in the mass of $D$ meson in medium is large enough then the higher charmonium states  may decay to $D\bar{D}$ pairs instead of $J/\psi$ states and this will further support the suppression of $J/\psi$ in HIC experiments. 
The drop in the mass of $D$ mesons in the medium will also decrease the threshold energy required for dissociation process like $J/\psi$ + $\pi$ $\to$ $D$ + $\bar{D}$
and will effect the
 absorption cross-section\cite{cross,absorp}.  
% The drop in the mass of $D$ mesons at finite density and temperature of the medium is observed in the mean-field models, QCD sum rules or quark meson coupling model calculations.
%are considered as major source of $J/\psi$ states and  decrease in the mass of $D$ meson in hadronic  environment may cause the decay of  higher charmonium states to.
On the contrary, if the mass of $D$ meson increase in the medium,
as was observed in PNJL model calculations, then  these mesons may act as facilitators to the production of $J/\psi$ state in the HIC experiments \cite{blaschke1}. 
% Therefore, to avoid mistakenly consideration of hadronic phase as the QGP phase, the in-medium study of $D$ meson become obligatory.

%   Additionally, the in-medium $D$ meson mass will be useful in  calculation of in-medium absorption cross-section for the process like $J/\psi$ + $\pi$ $\to$ $D$ + $\bar{D}$,  \cite{cross}. Particularly, the addition of scalar $D_0$ meson, may also have significant effect on the measurement of the absorption cross-section of $J/\psi$ by $\pi$ meson \cite{absorp}. 
   
     Moreover, the study of  $D$ mesons in nuclear as well as in strange hadronic matter might enlight the formation of bound state of $D$ meson with nucleons \cite{tsu} as well as with hyperons \cite{hoff}.   Further, the calculation of in-medium ratios of decay constants, $\frac{f_{D_{s0}}}{f_{D_0}}$, of charmed scalar mesons may also be used to measure the extent of flavour symmetry breaking in the strange hadronic matter as is done for pseudoscalar D mesons, $\frac{f_{D_{s}}}{f_D}$ \cite{domi2,nari,baza}. Upcoming experiment of  Facility for Antiproton and Ion Research (FAIR) project at GSI, Germany will    provide an unique opportunity to study the in medium effects on the open and hidden charmed mesons.   The Compressed Baryonic Matter (CBM) and anti-Proton Annihilation at Darmstadt (PANDA) will focus on the charmed spectroscopy and on in-medium decay widths of the charmed hadrons. CBM experiment may explore the phase of high baryonic density and moderate temperature which is just complement to the Relativistic Heavy Ion Collider (RHIC) and Large Hadron Collider (LHC). Apart from this, open charmed mesons are expected to be produced in the J-PARC facility which motivate us to study the properties of $D$ mesons in nuclear as well as in strange hadronic matter \cite{babar,J}. 
     The study of in-medium behavior of $D$ mesons may help to understand the experimentally observed elliptic flow, $v_2$, and nuclear modification factor, $R_{AA}$, of these meson, as described in recent review \cite{review_compare}.
     
%     As $D$ meson contain ($c\bar{q}$), where $c$ is heavy charm quark and $q$ is any light quark ($u$, $d$ or $s$), and we know that, in primordial heavy ion collisions, heavy quarks are expected to be produced and during their entire evolution in the medium, these heavy quarks can explore the properties of the created medium. Further, during hadronization, $c$ quark forms bound state with light quark $q$, and final $D$ mesons state can be detected. 
%      In this point, two interesting observables are the 
%      The study of in-medium behavior of $D$ mesons also help to understand the experimentally observed elliptic flow and nuclear modification factor, $R_{AA}$, of these meson, as described in recent review \cite{review_compare}. Although, these are the experimental observables, many phenomenological models have also been developed to reproduce the experimental results as described in recent review \cite{review_compare}. In this sense, the in-medium mass of $D$ meson may be helpful in precise theoretical measurement of elliptic flow, $v_2$, and nuclear modiication factor,  $R_{AA}$. 

On the phenomenological side, many methodologies have been developed to study the in-medium properties of $D$ mesons.  For example, Quark Meson Coupling Model (QMC) had predicted a negative shift in the  mass of $D$ meson \cite{tsu}. The self-consistent coupled channel approach predicted a positive and negative shift in the mass for pseudoscalar $D$ \cite{tolo1} and $D_s$ meson \cite{ CE}, respectively. This model was also used to investigate the scalar charm resonances of $D_{s0}(2317)$ and $D_0(2400)$ mesons \cite{tolo827} and observed the large medium effects for $D_{s0}(2317)$ meson as compared to $D_0(2400)$ meson. Here the QMC model treat  the quarks and gluons as degrees of freedom, and interactions between $D$ mesons and nucleons are taken through the exchange of scalar and vector mesons. On the other hand, self consistent coupled channel approach  considers the hadrons as degrees of freedom \cite{tolo}, and this undergo necessary modifications e.g., from SU(3) flavour \cite{tolo}, to SU(4) and breaking of SU(4) symmetry via exchange of vector mesons \cite{hoff, lutz1}. 

Another approach is QCD sum rules, in which the operator product expension (OPE) is applied on the current-current correlation function \cite{azizi}. In this analysis, using the Borel transformation the mass dependent terms are related with the quark as well as gluon condensates \cite{haya2000,hilger2009}. 
The properties of the scalar $D_0$ mesons in nuclear medium have also been studied  using QCD sum rule analysis upto leading order term \cite{wangs} and upto next to leading order term \cite{wang2}. In this technique the quark and gluon condensates needed for the QCD sum rules analysis were calculated using linear density approximation. The chiral SU(3) model generalized to SU(4) sector, had also been used to investigate the shift in the  masses of $D$ mesons \cite{arvinds2,pathaks,su4,chiral41}.
In \cite{arvinds}, the  chiral SU(3) model in-conjunction with QCD sum rules was applied successfully to study the in-medium masses of scalar mesons in nuclear medium. The in-medium properties of  pseudoscalar, vector and axial vector $D$ meson were investigated in \cite{Rahul,rahul2}.
In the present work, we will evaluate the shift in the masses and decay constants of scalar $D_0$ and  $D_{s0}$ mesons in asymmetric strange hadronic medium at finite temperatures. The in-medium properties of scalar $D_{s0}$ mesons were not addressed in \cite{arvinds} and owing to the presence of strange quark, the behaviour of these mesons in strange matter will of be considerable interest.

Furthermore, as an application of our work we shall investigate the in-medium partial decay width of $D_0$ and $D_{s0}$ for process  $D_0$ $\to$ $D+\pi$ ($D_{s0}$ $\to$ $D_s+\pi$). To achieve this goal, we use $^3P_0$ model \cite{micu}, 
which has been widely used in the past to evaluate the two body decay of the various mesons
 \cite{micu,yao,barn1,barn2,close,sego,ferre1,ferre2,ferre3,close1,zhong,chen,Li1}. The medium effects will be introduced through the medium modified  mass of these mesons. Here, we use  the in-medium mass of pseudoscalar $D$ meson as calculated in our previous work using chiral SU(3) model and QCD sum rule approach \cite{rahul2}.
 Additionally we take the in-medium pion mass as calculated using the in-medium chiral perturbative theory \cite{pionmass}

 This article is organized as follows:  In \cref{chiral}, we briefly describe the chiral SU(3) model to calculate in-medium  quark and gluon condensates. The QCD sum rules used to investigate the in-medium masses and decay constants  of $D_0$ and  $D_{s0}$ mesons is  discussed in \cref{QCD}, while the $^3P_0$ model used to evaluate in-medium partial decay width of $D_0(D_{s0})$ mesons is narrated in \cref{3P0}. In \cref{results}, we present the various results of the present work and finally in \cref{summary}, we shall summarize the present work. 
  \section{Chiral SU(3) model}
  \label{chiral}
We use the chiral SU(3) model to calculate the in-medium values 
of light quark condensates ($\left\langle \bar{u}u\right\rangle_{\rho_{B}}$, $\left\langle \bar{d}d\right\rangle_{\rho_{B}}$), strange quark condensates $\left\langle \bar{s}s\right\rangle_{\rho_{B}}$ and gluon condensates $\left\langle  \frac{\alpha_{s}}{\pi} {G^a}_{\mu\nu} {G^a}^{\mu\nu}
\right\rangle_{\rho_{B}}$. 
%The chiral SU(3) model is an effective model, based on the spontaneous breaking properties, broken scale invariance and nonlinear realization properties of chiral symmetry \cite{papa_nuclei, arvind2}. 
 Chiral SU(3) model contains an effective Lagrangian density, which include kinetic energy term, baryon meson interaction term which produce baryon mass, self-interaction of vector mesons which generates the dynamical mass of vector mesons, scalar mesons interactions which induce the spontaneous  breaking of chiral symmetry, and the explicit breaking term of chiral symmetry.
In the strange hadronic medium, in-medium baryon masses are modified in chiral SU(3) model through the exchange of scalar iso-scalar mesons $\sigma$ and $\zeta$ and scalar iso-vector field $\delta$.
Within mean field approximation,  from the effective Lagrangian density of the model, using Euler Lagrange equation $\frac{\partial \mathcal{L}}{\partial \phi}-\partial_\mu (\frac{\partial \mathcal{L}}{\partial(\partial_\mu \phi)})=0$, where $\phi$ is  scalar field, we obtain equations of motion for $\sigma$, $\zeta$, $\delta$ and scalar dilaton field $\chi$. These are given as \cite{ss,arvind2}
\begin{align}
& k_{0}\chi^{2}\sigma-4k_{1}\left( \sigma^{2}+\zeta^{2}
+\delta^{2}\right)\sigma-2k_{2}\left( \sigma^{3}+3\sigma\delta^{2}\right)
-2k_{3}\chi\sigma\zeta \nonumber\\
-&\frac{d}{3} \chi^{4} \bigg (\frac{2\sigma}{\sigma^{2}-\delta^{2}}\bigg )
+\left( \frac{\chi}{\chi_{0}}\right) ^{2}m_{\pi}^{2}f_{\pi}
-\sum g_{\sigma i}\rho_{i}^{s} = 0,
\label{sigma}
\end{align}
\begin{align}
& k_{0}\chi^{2}\zeta-4k_{1}\left( \sigma^{2}+\zeta^{2}+\delta^{2}\right)
\zeta-4k_{2}\zeta^{3}-k_{3}\chi\left( \sigma^{2}-\delta^{2}\right)\nonumber\\
-&\frac{d}{3}\frac{\chi^{4}}{\zeta}+\left(\frac{\chi}{\chi_{0}} \right) 
^{2}\left[ \sqrt{2}m_{K}^{2}f_{K}-\frac{1}{\sqrt{2}} m_{\pi}^{2}f_{\pi}\right]
 -\sum g_{\zeta i}\rho_{i}^{s} = 0,
\label{zeta}
\end{align}
\begin{align}
 & k_{0}\chi^{2}\delta-4k_{1}\left( \sigma^{2}+\zeta^{2}+\delta^{2}\right)
\delta-2k_{2}\left( \delta^{3}+3\sigma^{2}\delta\right) +k_{3}\chi\delta 
\zeta \nonumber\\
 + &  \frac{2}{3} d \chi^4 \left( \frac{\delta}{\sigma^{2}-\delta^{2}}\right)
-\sum g_{\delta i}\rho_{i}^{s} = 0,
\label{delta}
\end{align}
 
\begin{align}
 & k_{0}\chi \left( \sigma^{2}+\zeta^{2}+\delta^{2}\right)-k_{3}
\left( \sigma^{2}-\delta^{2}\right)\zeta + \chi^{3}\left[1
+{\rm {ln}}\left( \frac{\chi^{4}}{\chi_{0}^{4}}\right)  \right]
+(4k_{4}-d)\chi^{3}
\nonumber\\
 - & \frac{4}{3} d \chi^{3} {\rm {ln}} \Bigg ( \bigg (\frac{\left( \sigma^{2}
-\delta^{2}\right) \zeta}{\sigma_{0}^{2}\zeta_{0}} \bigg ) 
\bigg (\frac{\chi}{\chi_0}\bigg)^3 \Bigg )\nonumber\\ 
 + &\frac{2\chi}{\chi_{0}^{2}}\left[ m_{\pi}^{2} f_{\pi}\sigma +\left(\sqrt{2}m_{K}^{2}f_{K}-\frac{1}{\sqrt{2}}
m_{\pi}^{2}f_{\pi} \right) \zeta\right]  = 0,
\label{chi}
\end{align}
respectively.
In above, $m_\pi$, $m_K$ and $f_\pi$, $f_K$ denote the mass and  decay constant of $\pi$, $K$ mesons, respectively and  the other parameters $k_0, k_1, k_2$, $k_3$ and $k_4$ are fitted  so as to reproduce the vacuum masses of $\eta$ and  $\eta'$  mesons \cite{papa_nuclei}. Further, ${\rho_i}^s$ represents the scalar density for $i^{th}$ baryon ($i=p, n,$ $\L$, $\Sigma^{\pm,0}$, $\Xi^{-,0}$) and is defined as 
\begin{align}
\rho_{i}^{s} = \gamma_{i}\int\frac{d^{3}k}{(2\pi)^{3}} 
\frac{m_{i}^{*}}{E_{i}^{*}(k)} 
\Bigg ( \frac {1}{e^{({E_i}^* (k) -{\mu_i}^*)/T}+1}
+ \frac {1}{e^{({E_i}^* (k) +{\mu_i}^*)/T}+1} \Bigg ),
\label{scaldens}
\end{align}
where, ${E_i}^*(k)=(k^2+{{m_i}^*}^2)^{1/2}$ and ${\mu _i}^* 
=\mu_i -g_{\omega i}\omega -g_{\rho i}\rho -g_{\phi i}\phi$, are the single 
particle energy and the effective chemical potential
for the baryon of species $i$, and
$\gamma_i$=2 is the spin degeneracy factor. Also, $m_i^*$ = $- g_{\sigma i} \sigma - g_{\zeta i} \zeta - g_{\delta i}\delta$ is the effective mass of the baryons in the asymmetric hadronic medium. Parameters $g_{\sigma i}$, $g_{\zeta i}$ and $g_{\delta i}$ are fitted to reproduce the vacuum baryon masses \cite{papa_nuclei}. In \cref{chi} $\sigma_0$, $\zeta_0$ and $\chi_0$ denote the vacuum values of the scalar fields $\sigma$, $\zeta$ and $\chi$, respectively.

%For given density $\rho_B$ of the baryonic medium, we solve  coupled equations of motion of scalar fields using mean field approximation for the different values of strangeness fractions $f_s$, isospin asymmetric parameter $I$ and temperature $T$. The strangeness fraction is defined as $f_s$ = $\frac{\Sigma_i |s_i|\rho_i}{\rho_B}$, here $s_i$ is the number of strange quarks and $\rho_i$ is number density of $i^{th}$ baryon defined by $\rho_{i} = \gamma_{i}\int\frac{d^{3}k}{(2\pi)^{3}} 
% ( \frac {1}{e^{({E_i}^* (k) -{\mu_i}^*)/T}+1}
%+ \frac {1}{e^{({E_i}^* (k) +{\mu_i}^*)/T}+1})$. Further, the isospin asymmetric parameter is defined as $I = -\frac{\Sigma_i I_{3i} \rho_i}{2\rho_B}$, where $I_{3i}$ is the $z$-component of  the isospin for the $i^{\text{th}}$ baryon \cite{arvind2}.
 Furthermore, we solve these equations to find the effect of baryonic density ($\rho_{B}$), temperature ($T$), finite strangeness fraction $(f_s=\frac{\Sigma_i |s|_i \rho_i}{\rho_{B}})$ and isospin asymmetric parameter ($I = -\frac{\Sigma_i I_{3i} \rho_i}{2\rho_{B}}$) on $\sigma$, $\zeta$, $\delta$ and $\chi$ fields. Here, it is to be noted that, $I_{3i}$ is the $z$-component of  the isospin for the $i^{\text{th}}$ baryon, $s_i$ is the number of strange quarks and $\rho_i$ is the number density of $i^{th}$ baryon.

%In the present work to calculate the in-medium properties of $D_0$ and $D_{s0}$ meson in asymmetric medium, we shall need to evaluate the in-medium values of scalar condensates $\left\langle \bar{u}u\right\rangle_{\rho_{B}}$, $\left\langle \bar{d}d\right\rangle_{\rho_{B}}$ and of strange quark condensates $\left\langle \bar{s}s\right\rangle_{\rho_{B}}$. 
In the chiral SU(3) model, the explicit symmetry breaking term is used to relate the light and strange quark condensates with $\sigma$, $\zeta$, $\delta$ and $\chi$ fields as follows \cite{ss},

 \begin{align}
\left\langle \bar{u}u\right\rangle
= \frac{1}{m_{u}}\left( \frac {\chi}{\chi_{0}}\right)^{2}
\left[ \frac{1}{2} m_{\pi}^{2}
f_{\pi} \left( \sigma + \delta \right) \right],
\label{qu}
\end{align}

\begin{align}
\left\langle \bar{d}d\right\rangle
= \frac{1}{m_{d}}\left( \frac {\chi}{\chi_{0}}\right)^{2}
\left[ \frac{1}{2} m_{\pi}^{2}
f_{\pi} \left( \sigma - \delta \right) \right],
\label{qd}
\end{align}

and 
\begin{align}
\left\langle \bar{s}s\right\rangle
= \frac{1}{m_{s}}\left( \frac{\chi}{\chi_0} \right)^2\left( \sqrt{2}m_K^2 f_K  - \frac{1}{\sqrt{2}} m_ {\pi}^2 f_{\pi}\right) \zeta,
\label{qs}
\end{align}
respectively.

%Here $m_s$, $m_u$ and $m_d$ denotes the mass of $s$ quark, $u$ quark and $d$ quark respectively. Further the term ($m_K$, $f_K$) and ($m_{\pi}$, $f_{\pi}$) denote the (mass, decay constant) of kaon and pion, respectively.
  Furthermore, using the trace anomaly property of QCD we extract the gluon condensates in terms of the above mentioned scalar fields using \cite{papa_nuclei,arvind2}

\begin{align}
\left\langle  \frac{\alpha_{s}}{\pi} {G^a}_{\mu\nu} {G^a}^{\mu\nu}
\right\rangle =  \frac{8}{9} \Bigg [(1 - d) \chi^{4}
+\left( \frac {\chi}{\chi_{0}}\right)^{2}
\left( m_{\pi}^{2} f_{\pi} \sigma
+ \big( \sqrt {2} m_{K}^{2}f_{K} - \frac {1}{\sqrt {2}}
m_{\pi}^{2} f_{\pi} \big) \zeta \right) \Bigg ].
\label{glu}
\end{align}
In the above equation, $d$ denotes a constant with a value of $(2/11)$, and it is evaluated using one loop beta function for the three flavors and colors of QCD \cite{papa_nuclei}.  

%Initially effective model based on the $SU(2)_L * SU(2)_R$ symmetry and scale invariance was introduced and successfully applied to consider the effect of finite temperature on the nuclear matter \cite{heide, papa55}. 
  
  \section{QCD sum rule for $D_0$ and $D_{s0}$ mesons}
  \label{QCD}
We will now present the QCD sum rules to investigate the in-medium masses and decay constants of $D_0$ and $D_{s0}$ mesons. In doing so, one start with two point correlation function
 \begin{align}
\Pi (q) = i\int d^{4}x\ e^{iq.x} \langle \mathcal{T}\left\{J(x)J^{\dag}(0)\right\} \rangle_{\rho_{B}, T} ,
\label{tw}
\end{align}
where $\mathcal{T}$ is the time-ordered covariant operator and in the present work this will act on the scalar currents for the   $D_0$ and $D_{s0}$  mesons, given as \cite{arvinds}
\begin{align}
 J(x) &= J^\dag(x) =\frac{\bar{c}(x) q(x)+\bar{q}(x) c(x)}{2}.
 \label{psc}
 \end{align}
Note that in above, we consider the averaged scalar currents of particle $D_0$ and its antiparticle $\bar{D}_0$ mesons and thus, we will evaluate the averaged shift in masses and decay constants of scalar $D_0$  and similarly, for $D_{s0}$ mesons 
\cite{koi,wang1,azizi}. 
As mentioned earlier, we will evaluate the properties of $D_0$ and $D_{s0}$ meson in isospin asymmetric matter. The finite isospin asymmetry of the medium will cause the splitting in masses of $D_0^+$ and $D_0^0$  mesons belonging to isospin doublet of scalar $D_0$ mesons. In \cref{psc}, for $D_0^+$ and $D_0^0$  mesons quark field $q(x)$ will be replaced by $d(x)$ and $u(x)$, respectively, whereas for $D_{s0}$ mesons $q(x)$ will be replaced by $s(x)$. 
 The mass splitting between particle-antiparticles can be evaluated
 by separating the two point correlation function into even and odd part as was done in \cite{hilger277}. 
%On the other hand, in \cite{hilger277}, authors observed the quantitative splitting between $D_0$ and $\bar{D_0}$ mesons by separating the two point correlation function into even and odd part. 
In the rest frame of nucleons, following the Fermi gas approximation, we divide the two point  correlation function into vacuum part, nucleon and temperature dependent part, i.e.,

%\Pi (q) &=&\Pi_{0} (q)+ \frac{\rho_{D}}{2m_N}T_{N} (q) + \Pi_{P.B.}(q)\,,
%\label{pib}
% \end{eqnarray}
% 
 \begin{align}
\Pi (q) =\Pi_{0} (q)+ \frac{\rho_{B}}{2m_N}T_{N} (q) + \Pi_{P.B.}(q,T)\,,
\label{pb}
 \end{align}

 where $T_N (q)$ is the forward scattering amplitude, $\rho_{B}$ and $m_N$ denote the total baryon density and nucleon mass, respectively. The third term represents the thermal correlation function and is defined as \cite{Elet}
\begin{align} \label{pb2}
\Pi_{P.B.}(q, T) = i\int d^{4}x\ e^{iq. x} \langle \mathcal{T}\left\{J(x)J^\dag(0)\right\} \rangle_{T},
\end{align}
where $\langle \mathcal{T}\left\{J_5(x)J_5^\dag(0)\right\}\rangle_{T}$  is the thermal average of the time ordered product of the scalar currents.
Further, thermal average of any operator $\mathcal{O}$ is given by \cite{Elet}
%\begin{equation}
%\left\langle \mathcal{O} \left\rangle_T = \frac{$\text{Tr}$\left[ $\text{exp}$\left(-H/T\right) \mathcal{O} \right]}{$\text{Tr}$ $\text{exp}$\left( -H/T\right)}
%\end{equation}
\begin{align}
\left\langle \mathcal{O} \right\rangle _T = \frac{Tr \left\lbrace  \text{exp}\left(-H/T\right) \mathcal{O}\right\rbrace}{Tr \left\lbrace \text{exp}\left(-H/T\right)\right\rbrace}.
\end{align}
In above $Tr$ denotes the trace over complete set of states and $H$ is the QCD Hamiltonian. The factor $\frac{\text{exp}\left(-H/T\right)}{Tr \left\lbrace\text{exp}\left(-H/T\right) \right\rbrace}$  is the thermal density matrix of QCD. In \cref{pb}, the third term corresponds to the pion bath term and had been widely used in the past to consider the effect of temperature of the medium \cite{zsch,kwon}. 
Here we point out that, we consider the effect of temperature at finite baryonic density on the properties of $D_0$ and $D_{s0}$ mesons through the temperature dependence of scalar fields $\sigma$, $\zeta$, $\delta$ and $\chi$ in terms of which scalar quark and gluon condensates are expressed and therefore, we neglect the third term  in \cref{pb}.  The scattering amplitude $T_N (q)$, near the pole position of the scalar meson is represented in terms of the spectral density \cite{koi}, which in the limit of $\textbf{q}$ $\to$ 0, is parametrized in terms of three unknown parameters $a$, $b$ and $c$, given as    \cite{wang1,wang2,haya2000},
\begin{align}\label{a1}
\rho(\omega,0) &= -\frac{f_{D_0/D_{s0}}^2m_{D_0/D_{s0}}^4}{\pi m_{c}^2} \nonumber
 \mbox{Im} \left[\frac{{{T}_{D_0/D_{s0}}}(\omega,{\bf 0})}{(\omega^{2}-
m_{D_0/D_{s0}}^2+i\varepsilon)^{2}} \right] + \cdots  \\ \nonumber
\end{align}
%\begin{align} %\label{a1}
%\rho(\omega,0) &= -\frac{f_{D_s/B_s}^2m_{D_s/B_s}^4}{\pi m_{c/b}^2} \nonumber
% \mbox{Im} \left[\frac{{{T}_{D_s/B_s}}(\omega,{\bf 0})}{(\omega^{2}-
%m_{D_s/B_s}^2+i\varepsilon)^{2}} \right] + \cdots  \\ \nonumber
%\label{a1}
%\end{align}
\begin{align}
& = a\,\frac{d}{d\omega^2}\delta(\omega^{2}-m_{D_0/D_{s0}}^2)
 +
b\,\delta(\omega^{2}-m_{D_0/D_{s0}}^2) + c\,\theta(\omega^{2}-s_{0})\,.
\label{a2}
\end{align}

Here,  $m_{D_0/D_{s0}}$ and $f_{D_0/D_{s0}}$ are the masses and decay constants of $D_0/D_{s0}$ mesons and $m_c$ denotes the mass of charm quark. Also the first term in \cref{a2}  denotes the double pole term and exhibit the on shell effect of the $T$-matrix, whereas the second term represents single pole term which exhibit the off-shell effect of the $T$-matrix.  
%The second term corresponds to the off-shell  effects of the $T$-matrix, 
%and can be used to find the equation 
%%  \begin{align}\label{Tmatrix}
%%T_N(\omega,0)=\int^{+\infty}_{-\infty}du\frac{\rho(u,0)}{u-\omega-i\varepsilon}=\int_{0}^{\infty}du^2\frac{\rho(u,0)}{u^2-\omega^2},
%%\end{align}
%\begin{align}\label{ab}
%a&=-\frac{8\pi
%f_{D_0/D_{s0}}^2m_{D_0/D_{s0}}^4(M_N+m_{D_0/D_{s0}})}{m^2_{c/b}}a_{D_0/D_{s0}},
%\end{align}  
% here $a_{D_0/D_{s0}}$ is the scattering length of the corresponding meson.
 The third term proportional to $c$, corresponds to the continuum  term.  Here, $s_0$ is the continuum threshold parameter, and its value is fixed to reproduce the vacuum masses for $D_0$ and $D_{s0}$ mesons \cite{wangs}. 
 %In this manner, through \cref{a2,pb,pb2} we find the 
 Finally, the shift in masses and decay constants of $D_0$/$D_{s0}$ mesons from their vacuum value is given as \cite{arvinds,wangs}    
    
    \begin{equation}
\delta m_{D_0/D_{s0}} = 2\pi \frac{m_N + m_{D_0/D_{s0}}}{m_N m_{D_0/D_{s0}}} \rho_{B} a_{D_0/D_{s0}},
\label{masshift}
\end{equation}
  and 
 \begin{equation}
 \delta f_{D_0/D_{s0}} =  \frac{m_{c} ^2}{2f_{D_0/D_{s0}} m^4}\left(\frac{b \rho_B}{2m_N} - \frac{4 f_{D_0/D_{s0}}^2 m_{D_0/D_{s0}}^3 \delta m_{D_0/D_{s0}}}{m_{c} ^2}\right),
 \label{decayshift}
\end{equation} 
respectively.
Clearly, in order to calculate the shift in mass and decay constant of $D_0/D_{s0}$ mesons we shall need to find the values of unknown parameters $a$ and $b$. To achieve this task, we apply the Borel transformation on the forward scatterring amplitude $T_N(\omega,0)$ on hadronic side as well as on  the forward scattering amplitude $T_N(\omega,0)$ on operator product expansion (OPE) side in the rest frame of the nuclear matter. After this, we equate these two equations and this lead to \cite{arvinds,wangs}
 
\begin{align}
& a \left\{\frac{1}{M^2}\exp\left(-\frac{m_{{D_0/D_{s0}}}^2}{M^2}\right) - \frac{s_0}{m_{{D_0/D_{s0}}}^4} \exp\left(-\frac{s_0}{M^2}\right)\right\} \nonumber\\
+& b \left\{\exp\left(-\frac{m_{{D_0/D_{s0}}}^2}{M^2}\right) - \frac{s_0}{m_{{D_0/D_{s0}}}^2} \exp\left(-\frac{s_0}{M^2}\right)\right\}\nonumber\\
+&  \frac{2m_N(m_H-m_N)}{(m_H-m_N)^2-m_{{D_0/D_{s0}}}^2}\left(\frac{f_{{D_0/D_{s0}}}m_{{D_0/D_{s0}}}g_{{D_0/D_{s0}}NH}}{m_{c}}\right)^2\nonumber\\
\times & \left\{ \left[\frac{1}{(m_H-m_N)^2-m_{{D_0/D_{s0}}}^2}-\frac{1}{M^2}\right] 
\exp\left(-\frac{m_{{D_0/D_{s0}}}^2}{M^2}\right)\right.\nonumber\\
-&\left.\frac{1}{(m_H-m_N)^2-m_{{D_0/D_{s0}}}^2}\exp\left(-\frac{(m_H-m_N)^2}{M^2}\right)\right\}\nonumber\\
=&+\frac{m_{c}\langle\bar{q}q\rangle_N}{2}\times \exp\left(- \frac{m_{c/b}^2}{M^2}\right) \nonumber\\
+&\frac{1}{2}\left\{-2\left(1-\frac{m_{c}^2}{M^2}\right)\langle q^\dag i D_0q\rangle_N +\frac{4m_{c}
}{M^2}\left(1-\frac{m_{c}^2}{2M^2}\right)\langle \bar{q} i D_0 i D_0q\rangle_N+\frac{1}{12}\langle\frac{\alpha_sGG}{\pi}\rangle_N\right\} \nonumber\\
\times &\exp\left(- \frac{m_{c}^2}{M^2}\right)\, .
\label{qcdsum}
\end{align}
Here, to find the values of two unkown parameters  $a$ and $b$ we differentiate above equation w.r.t.  $\frac{1}{M^2}$ to find another equation and then solve these two equations.  The nucleon expectation value of the various condensates appearing in  \cref{qcdsum} is written as \cite{hilger2009}  
\begin{equation}
\mathcal{O}_{N} = \left[ \mathcal{O}_{\rho_{B}}  - \mathcal{O}_{vacuum}\right] \frac{2m_N}{\rho_{B}}.
\label{condexp}
\end{equation}
Explicitly, the nucleon expectation values of light quark and gluon condensates are expressed as, 
\begin{equation}
{<u \bar{u}>}_{N} = \left[ {<u\bar{u}>}_{\rho_{B}}  - {<u \bar{u}>}_{vacuum}\right] \frac{2m_N}{\rho_B},
\label{ucondexp1}
\end{equation}

\begin{equation}
{<d \bar{d}>}_{N} = \left[ {<d\bar{d}>}_{\rho_{B}}  - {<d \bar{d}>}_{vacuum}\right] \frac{2m_N}{\rho_B},
\label{ucondexp1}
\end{equation}
and 
\begin{equation}
\left\langle  \frac{\alpha_{s}}{\pi} {G^a}_{\mu\nu} {G^a}^{\mu\nu} 
\right\rangle_{N} = \left[ \left\langle  \frac{\alpha_{s}}{\pi} {G^a}_{\mu\nu} {G^a}^{\mu\nu} \right\rangle_{\rho_B} -  \left\langle  \frac{\alpha_{s}}{\pi} {G^a}_{\mu\nu} {G^a}^{\mu\nu} \right\rangle_{vacuum} \right]\frac{2m_N}{\rho_B}.
\label{Gcondexp1}
\end{equation}
The condensates $\langle\bar{q}g_s\sigma Gq\rangle_{\rho_B}$  and 
$\langle \bar{q} i D_0 i D_0q\rangle_{\rho_B}$ appearing in Borel transformed QCD sum rule equation are expressed in terms of  light quark condensates and we write\cite{arvinds,hilger2009}
\begin{align}
\langle\bar{q}g_s\sigma Gq\rangle_{\rho_B} = \lambda^{2}\left\langle \bar{q}q \right\rangle_{\rho_{B}} + 3.0 GeV^{2}\rho_{B}.
\label{cond2}
\end{align}
and
\begin{align}
\langle \bar{q} i D_0 i D_0q\rangle_{\rho_B} + \frac{1}{8}\langle\bar{q}g_s\sigma Gq\rangle_{\rho_B} =  0.3 GeV^{2}\rho_{B}.
\label{cond3}
\end{align}
The condensate $\langle q^\dag i D_0q\rangle_N$ is not calculated in the chiral SU(3) model and we consider its value as calculated in linear density approximation for our calculations. 
We will use the values 0.18 GeV$^2$ $\rho_B$ and 0.018 GeV$^2$ $\rho_B$ for $\langle u^\dag i D_0u\rangle_N$ and $\langle s^\dag i D_0s\rangle_N$, respectively \cite{hilger_q0}. 
However, later on we will see that  $\langle q^\dag i D_0q\rangle_N$ does not effect significantly the in-medium properties of
$D_0$ and $D_{s0}$ mesons.  
%take its value for $D_0$($D_{s0}$) meson as 0.18 Gev$^2$ $\rho_B$ (0.018 Gev$^2$ $\rho_B$) \cite{hilger_q0}, and later on we will explain its insignificant effect on the result of the present work.  Thus, using these in-medium values of condensates we calculate the in-medium masses and decay constants of $D_0$/$D_{s0}$ mesons. 
  
  %Methodology before to apply for the $D$ and $B$ mesons, is expected to verify the well developed results of the light mesons($\rho$, $\omega$ etc). 
  
\section{$^3P_0$ model} 
\label{3P0}
To calculate the in-medium partial decay width of
$D_0$ $\to$ $D+\pi$ ($D_{s0}$ $\to$ $D_s+\pi$), we use $^3 P_0$ model, in which quark and anti-quark pair  is created in vacuum ($0^{++}$) \cite{micu,yao}.
% first proposed by Micu \cite{micu}, and then modified for the OZI allowed decay of mesons \cite{yao}.
  This model had been used in literature to find the strong decays of hidden charmed states \cite{barn1, barn2}, open charmed bottom states \cite{close, sego} as well as of  bottom mesons \cite{ferre1, ferre2, sego, ferre3}. In the present work of finding the two body decay of $D_0/D_{s0}$ mesons, we use the transition operator as taken in \cite{Y(4040)}, and find the helicity amplitude given by \cite{liu} 
\begin{widetext}
\begin{align}\label{eq:M}
 \mathcal{M}^{M_{J_{D_0} } M_{J_{D} } M_{J_{\pi} }} = \gamma  \sqrt {8E_{D_0} E_{D} E_{\pi} } \sum_{\substack{M_{L_{D_0} } ,M_{S_{D_0} } ,M_{L_{D} }, \\M_{S_{D} } ,M_{L_{\pi}} ,M_{S_{\pi} } ,m} }\langle {1m;1 - m}|{00} \rangle \nonumber \\
 \times \langle {L_{D_0} M_{L_{D_0} } S_{D_0} M_{S_{D_0} } }| {J_{D_0} M_{J_{D_0} } }\rangle \langle L_{D} M_{L_{D} } S_{D} M_{S_{D} }|J_{D} M_{J_{D} } \rangle\langle L_{\pi} M_{L_{\pi} } S_{\pi} M_{S_{\pi} }|J_{\pi} M_{J_{\pi} }\rangle \nonumber \\
  \times\langle\varphi _{D}^{13} \varphi _{\pi}^{24}|\varphi _{D_0}^{12}\varphi _0^{34} \rangle
\langle \chi _{S_{D} M_{S_{D} }}^{13} \chi _{S_{\pi} M_{S_{\pi} } }^{24}|\chi _{S_{D_0} M_{S_{D_0} } }^{12} \chi _{1 - m}^{34}\rangle I_{M_{L_{D} } ,M_{L_{\pi} } }^{M_{L_{D_0}} ,m} (\textbf{k}).
\end{align}
\end{widetext}
In above,  $E_{D_0}$= $m^*_{D_0}$, $E_{D}$ = $\sqrt{m_{D}^{*2} + K_{D}^2}$ and $E_{\pi}$ = $\sqrt{m_{\pi}^{*2} + K_{\pi}^2}$ represent the energies of respective mesons. Here $m^*_{D_0}$, $m^*_{D}$ and $m^*_{\pi}$ are the in-medium masses of $D_0$, $D$ and $\pi$ mesons, respectively.
 We then calculate the spin matrix elements $\langle  \chi _{S_{D} M_{S_{D} }}^{13} \chi _{S_{\pi} M_{S_{\pi} } }^{24}|\chi _{S_{D_0} M_{S_{D_0} } }^{12} \chi _{1 - m}^{34}\rangle$ in terms of the Wigner's 9j symbol, and the flavor matrix element $\langle\varphi _{D}^{13} \varphi _{\pi}^{24}|\varphi _{D_0}^{12}\varphi _0^{34} \rangle$ in terms of isospin of quarks as done in Refs. \cite{Y(4040), liu, yao}.
  In \cref{eq:M},
  $I_{M_{L_{D} } ,M_{L_{\pi} } }^{M_{L_{D_0}} ,m} (\textbf{k})$  represents the spatial integral and is expressed in terms of
  wave functions of parent and daughter mesons.
  We use simple harmonic oscillator type wave functions  defined by
%  for the particular decay ($D_0$ $\to$ $D$ + $\pi$), and we solve by taking the SHO type equation for the mesons as, 
\begin{align}
\psi_{nL{M_L}}= (-1)^n(-\iota)^L R^{L+\frac{3}{2}} \sqrt{\frac{2n!}{\Gamma(n+L+\frac{3}{2})}} \exp\Big{(}\frac{-R^2k^2}{2}\Big{)} L_n^{L+\frac{1}{2}}(R^2k^2) Y_{lm}(\bf{k}).
\label{eq:wave}
\end{align} 
 Here, $R$ is the radius of the meson,
 $L_n^{L+\frac{1}{2}}(R^2k^2)$  represents associate Laguerre polynomial and 
 $Y_{lm}(\bf {k})$ denotes the spherical harmonic function.

By taking these calculations in hand, and following the  Jacob-Wick formula we transform the helicity amplitude into partial wave amplitude as follows
%$$\sum_{n=1}^{\infty}$$
%\begin{align}\label{eq:L}
%%\end{split}
%\mathcal{M}^{JL} (D_0 \to D \pi) &= \frac{{\sqrt {2{L} + 1} }}{{2{ J_{D_0}} + 1}}\sum_{M_{J_{D}},M_{ J_{\pi}}} \langle{{ L}0{J} M_{J_{D_0}}} |{J_{D_0} M_{J_{D_0} } }\rangle \nonumber \\
%&\times \left\langle {{ J_{D}} M_{J_{D}} { J_{\pi}} M_{J_{\pi}}}\right|\left. {{J} M_{J_{D_0} } } \right\rangle M^{M_{J_{D_0} } M_{J_{D}} M_{J_{\pi} } } (\textbf{K}).
%%\end{split}
%\end{align}
\begin{align}\label{eq:L}
%\end{split}
\mathcal{M}^{JL} (D_0 \to D \pi) &= \gamma \frac{\sqrt{2E_{D_0}E_{D}E_{\pi}}}{6\sqrt{3}} [I_0 -2I_1],
%\end{split}
\end{align}
where, 
\begin{align}
 I_0&= - 4\frac{\sqrt{3}}{\pi^{5/4}}
\frac{R_{D_0}^{5/2}R_{D}^{3/2}R_{\pi}^{3/2}}
{(R_{D_0}^{2}+R_{D}^{2}+R_{\pi}^{2})^{5/2}}             \Bigg\{
1-{\mathbf{k}_D}^2 \frac{(2R_{D_0}^{2}+R_{D}^{2}+R_{\pi}^{2})(R_D^2+R_\pi^2)}{4(R_{D_0}^{2}+R_{D}^{2}+R_{\pi}^{2})}\Bigg\}\nonumber
\\ 
&\times
\exp\bigg[-\frac{\mathbf{k}_{D}^{2}R_{D_0}^{2}(R_{D}^{2}+R_{\pi}^{2})}{8(R_{D_0}^{2}+R_{D}^{2}+R_{\pi}^{2})}\bigg],
\label{I0}
\end{align}
and
\begin{align}
 I_1=  4\frac{\sqrt{3}}{\pi^{5/4}}
\frac{R_{D_0}^{5/2}R_{D}^{3/2}R_{\pi}^{3/2}}
{(R_{D_0}^{2}+R_{D}^{2}+R_{\pi}^{2})^{5/2}} 
\times
\exp\bigg[-\frac{\mathbf{k}_{D}^{2}R_{D_0}^{2}(R_{D}^{2}+R_{\pi}^{2})}{8(R_{D_0}^{2}+R_{D}^{2}+R_{\pi}^{2})}\bigg].
\label{I1}
\end{align}

We then finally calculate the decay width, using
\begin{align}
%\end{split}
\Gamma  = \pi^2 \frac{|{\mathbf{k}_{D}}|}{m_{A}^2}\sum_{JL} |{\mathcal{M}^{JL}}|^2,
\label{eq:G}
\end{align}
where, $\gamma$ is the strength of the pair creation in the vacuum and its value is taken as 6.74 \cite{liu}. Also, $|\mathbf{k}_{D}|$ represents the momentum of the $D$ and $\pi$ mesons in the rest mass frame of $D_0$ meson and is given by, 
\begin{align}
%\end{split}
 |\mathbf{k}_{D}|= \frac{{\sqrt {[m_{D_0}^{*2}  - (m^*_D  - m^*_{\pi} )^2 ][m_{D_0}^{*2}  - (m^*_D  + m^*_{\pi} )^2 ]} }}{{2m^*_{D_0} }}.
\label{eq:K}
\end{align}

 Here, for the decay   
 $D_{s0} \to D_s \pi$ the values for $D_0$ will be replaced by $D_{s0}$ and $D$ with $D_s$.  Thus, through the in-medium mass of $D_0$/$D_{s0}$, $D$/$D_s$ and $\pi$  mesons, the in-medium partial decay widths of the processes $D_0$ $\to$ $D$ $\pi$ and $D_{s0} \to D_s \pi$ can be calculated.

\section{Results and discussion}
\label{results}
This section will elaborate the results of the present investigation. We use, nuclear saturation density, $\rho_0$ = 0.15 fm$^{-3}$, the average values of coupling constants for scalar $D_0/D_{s0}$ mesons  $g_{{{D_0/D_{s0}}N\Lambda_{\pi}}}$ $\approx$ $g_{{{D_0/D_{s0}}N\Sigma_{\pi}}}$  $\approx$ 6.74, the values of continuum threshold parameter $s_0$, for $D_0^+$, $D_0^0$ and $D_{s0}$ mesons as 8, 8 and 7 GeV$^2$, respectively.  The vacuum values of masses  of $D_0^+$, $D_0^0$ and $D_{s0}$ mesons are taken as 2.355, 2.350 and 2.317 GeV, whereas the vacuum values of decay constants are taken to be 0.334, 0.334 and 0.333 GeV, respectively. 
We shall represent the shift in masses and decay constants of $D_0^+$, $D_0^0$ and $D_{s0}$ mesons as a function of squared Borel mass parameter, $M^2$. To find the shift in masses and decay constants of $D_0^+$, $D_0^0$ and $D_{s0}$ mesons we choose a proper Borel window within which the least variation in the masses and decay constants is observed. We choose the Borel window for $D_0$ and $D_{s0}$ mesons  as (5-9) GeV$^2$.

\begin{table}
\begin{tabular}{|l|l|l|l|l|l|l|l|l|l|}
\hline
& & \multicolumn{4}{c|}{I=0}    & \multicolumn{4}{c|}{I=0.5}   \\
\cline{3-10}
&$f_s$ & \multicolumn{2}{c|}{T=0} & \multicolumn{2}{c|}{T=100MeV}& \multicolumn{2}{c|}{T=0}& \multicolumn{2}{c|}{T=100MeV}\\
\cline{3-10}
&  &$\rho_0$&$4\rho_0$ &$\rho_0$  &$4\rho_0$ & $\rho_0$ &$4\rho_0$&$\rho_0$&$4\rho_0$ \\ \hline 
$\delta m_{D_0^0}$ & 0& 87& 162 &76&156&78&148&72 &143\\ \cline{2-10}
&0.5&103  &171  & 93 & 162 &87 &150 & 80 & 145 \\ \cline{1-10}
$\delta m_{D_0^+}$&0&64 & 125 & 58 &120 &68 &127 & 62 &123 \\  \cline{2-10}
&0.5&76 &129 &69 &123 & 84 &139 & 79&145 \\  
 \cline{1-10}
$\delta m_{D_{s0}}$&0&81 & 158 & 67 &140 &73 &140 & 66 &137 \\  \cline{2-10}
&0.5&113 &234 &101 &214 & 120 &252 &106&224 \\  \hline
$\delta {f_{D_0^0}}$& 0&-10&-19.4 &-9&-18.6&-9.2&-17.5&-8.3 &-17.1\\ \cline{2-10}
&0.5&-11  &-20  &-10.8 & -19.2 &-10.3 &-17.7 & -9.3 & -16.9 \\ \cline{1-10}
$\delta {f_{D_0^+}}$&0&-7.5 & -14.4 & -6.6 &-13.9 &-7.9 &-14.7 & -7.1 &-14.4 \\  \cline{2-10}
&0.5&-8.9 &-15 &-7.9 &-14.3 & -9.8 &-16.2 & -8.7&-15.4\\ 
\cline{1-10}
$\delta {f_{D_{s0}}}$&0&-7.7 & -14 & -6.3 &-12 &-7 &-12.5 & -6.3 &-12 \\  \cline{2-10}
&0.5&-11 &-21 &-9.6 &-19 & -11.5 &-22.7 & -10&-20.4 \\  \hline
\end{tabular}
\caption{In above, we tabulate the values of shift in masses and decay constants  of $D_0^0$, $D_0^+$ and $D_{s0}$ mesons (in units of MeV).}
\label{table:tbl}
\end{table}

\subsection{Shift in masses and decay constants}
\label{sub_mass}   
\begin{figure}
\centering
\includegraphics[width=14cm,,height=12cm]{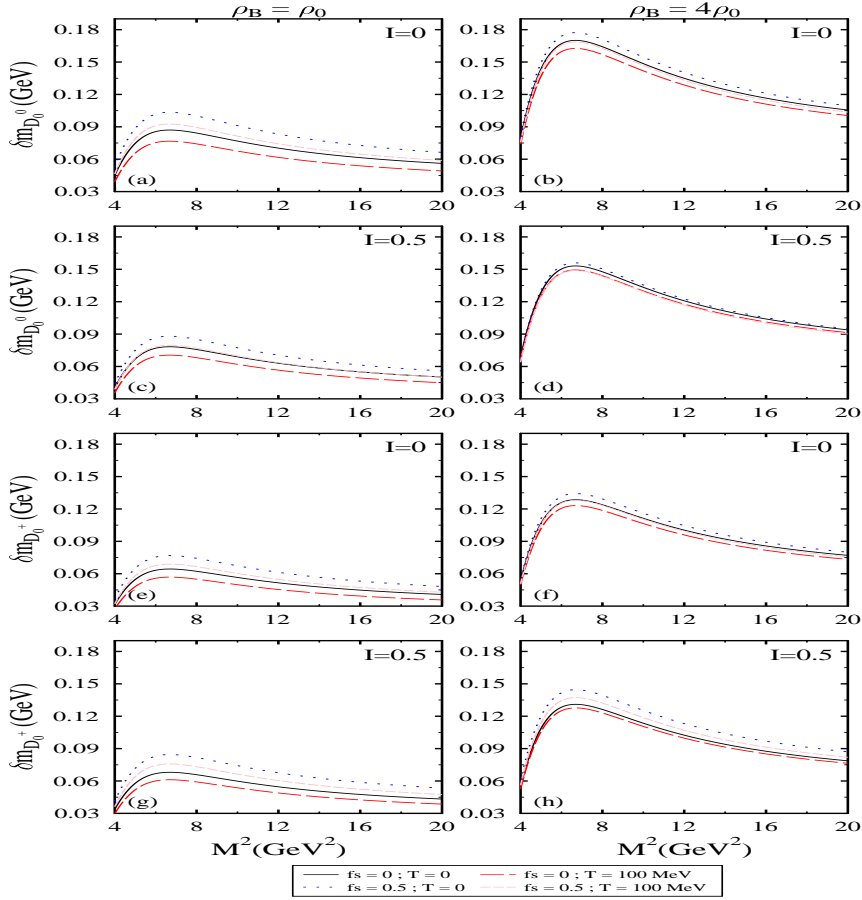}
\caption{Figure shows the variation of shift in masses of scalar $D_0^0$ and $D_0^+$ mesons as a function of squared Borel mass parameter, $M^2$ for  isospin asymmetric parameters $I = 0$ and $0.5$, temperatures $T = 0$ and 100 MeV and strangeness fractions $f_s = 0$ and $0.5$. The results are given at  baryonic densities $\rho_0$ and $4\rho_0$.}\label{fig:mass}
\end{figure}

\begin{figure}
\centering
\includegraphics[width=14cm,,height=12cm]{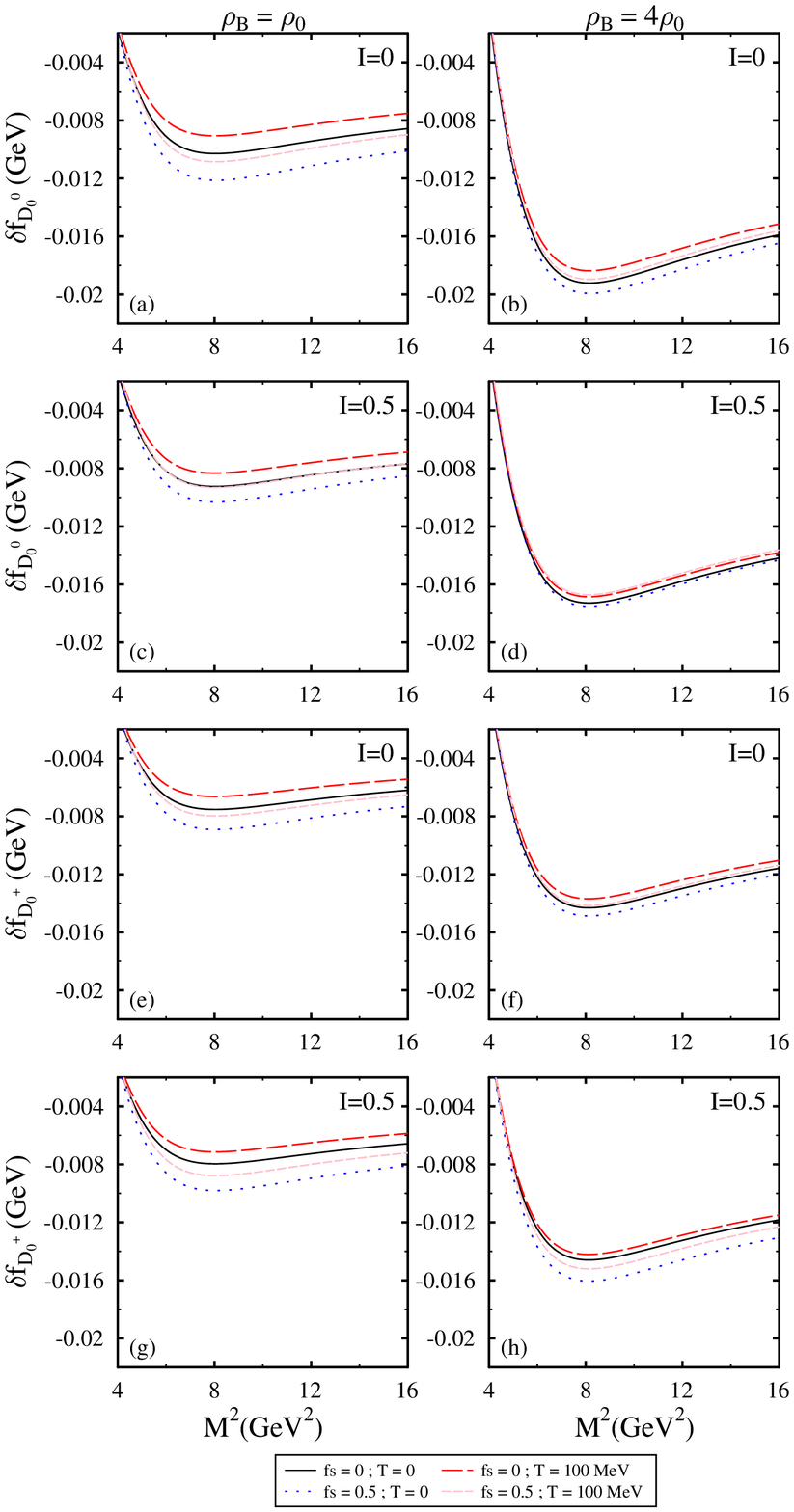}
\caption{Figure shows the variation of shift in decay constants of scalar $D_0^0$ and $D_0^+$ mesons as a function of squared Borel mass parameter, $M^2$ for isospin asymmetric parameters $I = 0$ and $0.5$, temperatures $T = 0$ and 100 MeV and strangeness fractions $f_s = 0$ and $0.5$. The results are given at  baryonic densities $\rho_0$ and $4\rho_0$.}\label{fig:decay}
\end{figure}

\begin{figure}
\centering
\includegraphics[width=14cm,,height=12cm]{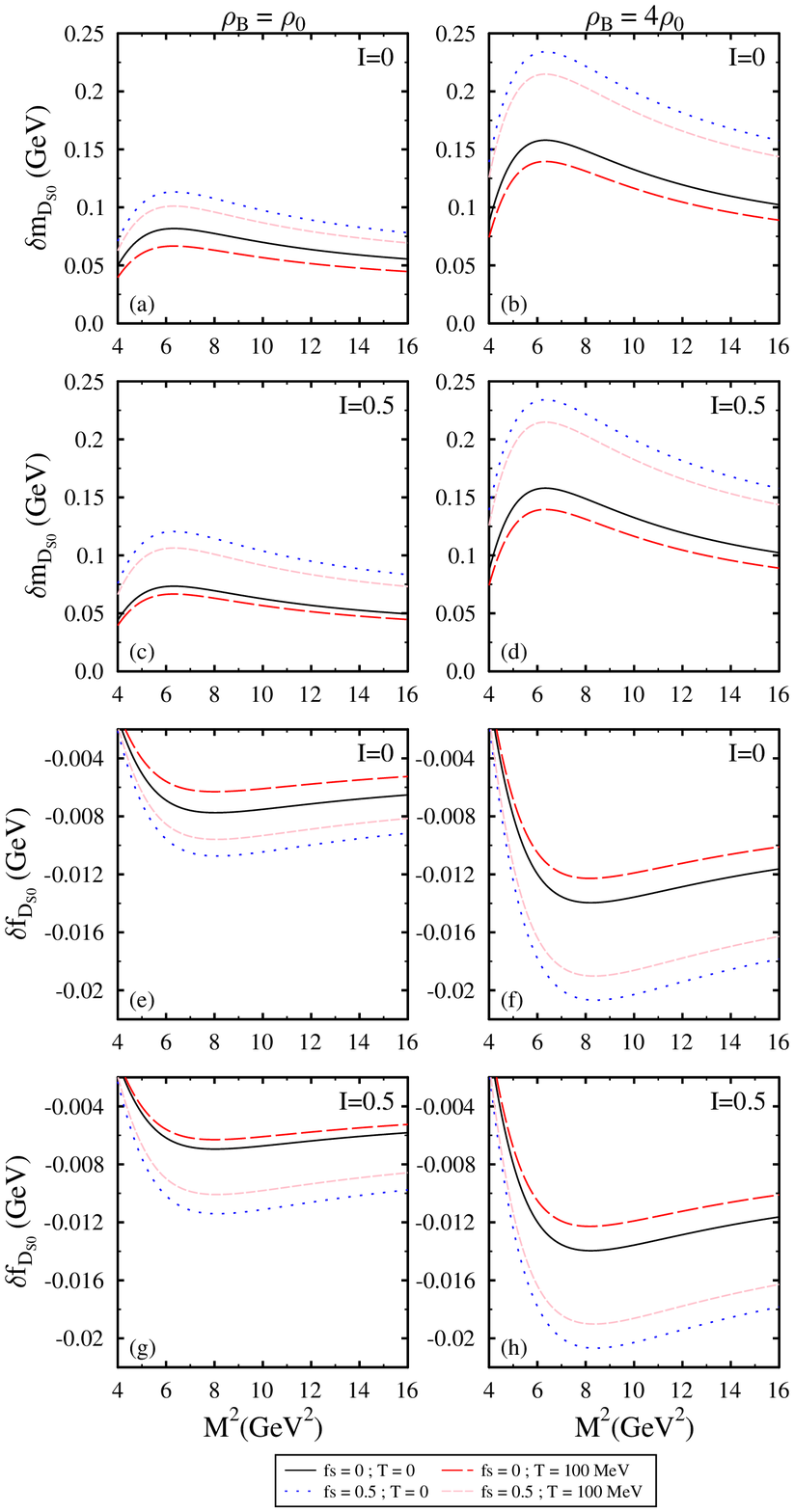}
\caption{Figure shows the variation of shift in mass and decay constant of scalar $D_{s0}$ mesons as a function of squared Borel mass parameter, $M^2$ for  isospin asymmetric parameters $I = 0$ and $0.5$, temperatures $T = 0$ and 100 MeV and strangeness fractions $f_s = 0$ and $0.5$. The results are given at  baryonic densities $\rho_0$ and $4\rho_0$.}\label{fig:massdecay}
\end{figure}

  In \cref{fig:mass} (\cref{fig:decay}) we represent the shift in masses (decay constants) of isospin doublet of scalar $D_0$ mesons, whereas in \cref{fig:massdecay} we plot the shift in masses and decay constants of $D_{s0}$ mesons in isospin asymmetric hot and dense strange hadronic medium as a function of squared Borel mass parameter, $M^2$. In \cref{table:tbl} we give the numerical values of shift in masses and decay constants of these mesons. 
 Here, in the present investigation, we notice an enhancement in the masses, whereas drop in the values of decay constants of scalar $D_0$ and $D_{s0}$ mesons in nuclear as well as in the strange hadronic matter. Moreover, for any given value  of isospin asymmetric parameter $I$, strangeness fraction $f_s$ and temperature $T$ of the medium the magnitude of the enhancement (drop) in the values of masses (decay constants) of $D_0$ and $D_{s0}$ mesons, increase as a function of baryonic density of the medium. 
For example, in symmetric nuclear medium, at temperature $T=0$ and baryonic density $\rho_B$=$\rho_0$, the masses (decay constants) of $D_0^0$, $D_0^+$ and $D_{s0}$ mesons increase (decrease) by 3.7$\%$ (2.9$\%$), 2.7$\%$ (2.2$\%$) and 3.5$\%$ (2.3$\%$), respectively from their vacuum values. Further, at baryonic density $4\rho_0$  of the same medium, the above values of percentage increase (decrease) change to 6.8$\%$ (5.8$\%$), 5.3$\%$ (4.3$\%$) and 6.8$\%$ (4.2$\%$), respectively.

% On the other hand, in symmetric nuclear medium the decay constants of these mesons decrease from their vacuum values by 2.9$\%$, 2.2$\%$ and 2.3$\%$, respectively $\rho_B$=$\rho_0$, $T$=0.
%Likewise, at higher baryonic density (4$\rho_B$) the above values change to  5.8$\%$, 4.3$\%$ and 4.2$\%$, respectively.

 Similar behaviour we observe for the shift in masses and decay constants of above mentioned mesons at finite strangeness fraction $f_s$. For example, in symmetric strange hadronic medium, $f_s$=0.5, the values of the masses (decay constants) of $D_0^0$, $D_0^+$ and $D_{s0}$ mesons increase (decrease) by 4$\%$ (3.2$\%$), 3$\%$ (2.6$\%$) and 4.8$\%$ (3.3$\%$), respectively from their vacuum values, at  $\rho_B$=$\rho_0$ and temperature $T$=0. Likewise, at baryonic density 4$\rho_0$,  these percentage values further enhance to  7$\%$ (5.9$\%$), 5.3$\%$(4.4$\%$) and 10$\%$(6.3$\%$), respectively. %from their vacuum values.
Further, we notice that the shift in masses and decay constants of $D_{s0}$ mesons is more sensitive to the finite strangeness fraction in the medium as compared to the non-strange $D_0$ mesons.
This can be understood on the basis that the in-medium mass and decay shift of $D_0$ mesons depend upon the light quark condensate $\left\langle \bar{q}q\right\rangle$, whereas that of $D_{s0}$ mesons is evaluated using strange quark condensates $\left\langle \bar{s}s\right\rangle$. As can be seen from \cref{qs}, the strange quark condensate $\left\langle \bar{s}s\right\rangle$ is proportional to the strange scalar field $\zeta$ which is more sensitive to the strangeness fraction of the medium as compared to non-strange scalar field $\sigma$.

%Also, the former is highly dependent on $\sigma$ field, whereas later one on $\zeta$ field as clear from \cref{qu,qd,qs}. Further, from chiral SU(3) model we note that, $\sigma$ field contain light $u,d$ quark and $\zeta$ contain strange $s$ quark.  Therefore, the properties of strange $D_{s0}$ meson are more sensitive to the presence of hyperons in the medium as compared to the non strange $D_0$ meson.
%  However, in zero temperature situations, and  baryonic density $\rho_B$, the decay constants of these mesons decrease from their vacuum values to 3.2$\%$, 2.3$\%$ and 3.2$\%$, respectively on shifting from nuclear to hyperonic medium (along with the nucleons) at $I$=0. 
  The effect of finite temperature on the mass and decay shift of above mentioned mesons is observed to be opposite to that of strangeness fraction. For example, at finite temperature medium i.e., $T=100$ MeV, we observe the percentage of increase (drop) in the masses (decay constants) of $D_0^0$, $D_0^+$ and $D_{s0}$ mesons as 6.7$\%$(5$\%$), 5.1$\%$(3.7$\%$) and 9$\%$(5.7$\%$), respectively from their vacuum values at $\rho_B$=4$\rho_0$, $f_s$=0.5 and $I$=0. Evidently,  these percentage values are lower than the  values  7$\%$(5.2$\%$), 5.3$\%$(3.9$\%$) and 10$\%$(6.2$\%$), respectively observed in the same medium but at zero temperature.   
%  Furthermore, in symmetric and strange hadronic matter, the  percentage drop in values of the decay constants of $D_0^0$, $D_0^+$ and $D_{s0}$ mesons are calculated as 5$\%$, 3.7$\%$ and 5.7$\%$, respectively at $T$=100 MeV, $\rho_B$=4$\rho_0$. Again, the magnitudes of these percentage change   are less than the values 5.2$\%$, 3.9$\%$ and 6.2$\%$, respectively observed at zero temperature situation and keeping the other parameters of the medium constant.   
Therefore, finite temperature of the medium cause decrease in the masses, whereas increase in the values of decay constants of  $D_0^0$, $D_0^+$ and $D_{s0}$ mesons.

\begin{figure}
\centering
\includegraphics[width=14cm,height=12cm]{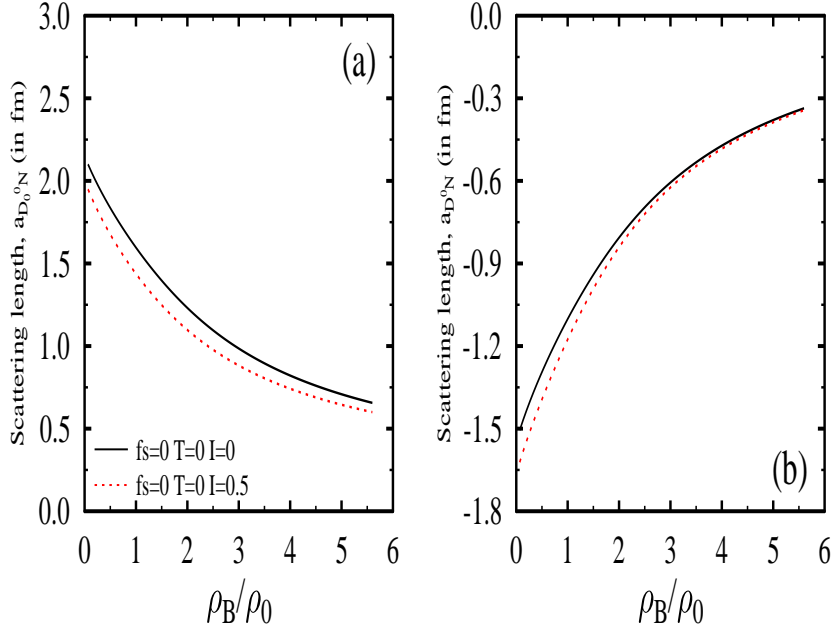}
\caption{Figure shows the variation of the scattering length(in fm) of scalar and pseudoscalar $D$ meson with nucleon in nuclear medium. }\label{fig:sle}
\end{figure}
\begin{figure}
\centering
\includegraphics[width=14cm,,height=12cm]{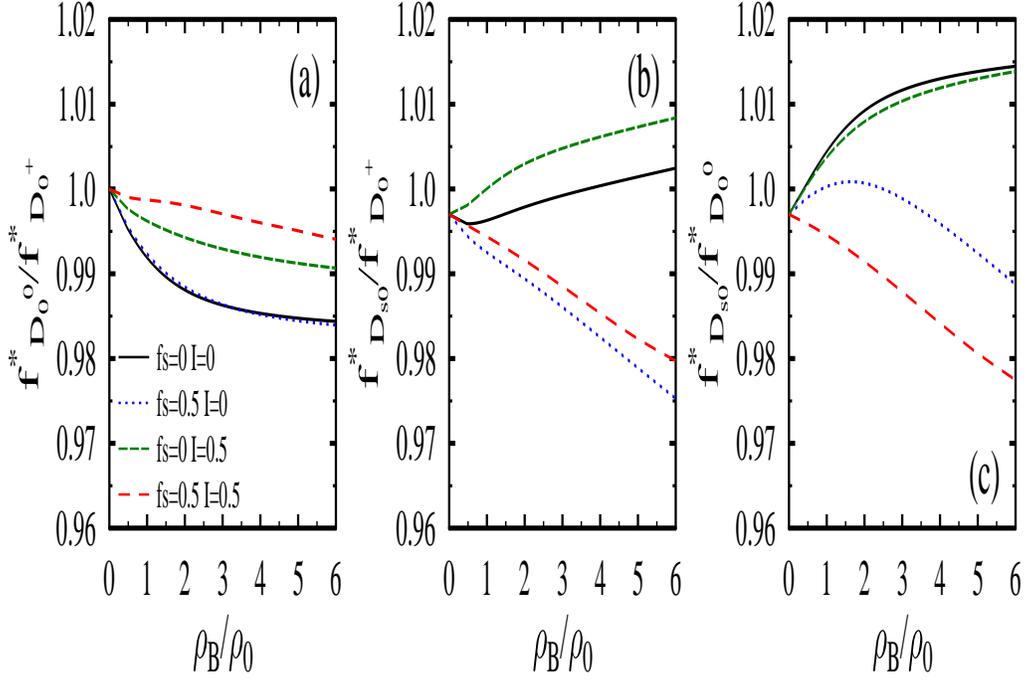}
\caption{Figure shows the variation of ratio of in-medium decay constants $\frac{f^*_{D^0_0}}{f^*_{D^+_0}}$, $\frac{f^*_{D_{s0}}}{f^*_{D^+}}$  and $\frac{f^*_{D_{s0}}}{f^*_{D^0}}$ as a function of baryonic density of the medium.}\label{fig:ratiod}
\end{figure}
\begin{figure}
\centering
\includegraphics[width=14cm,,height=12cm]{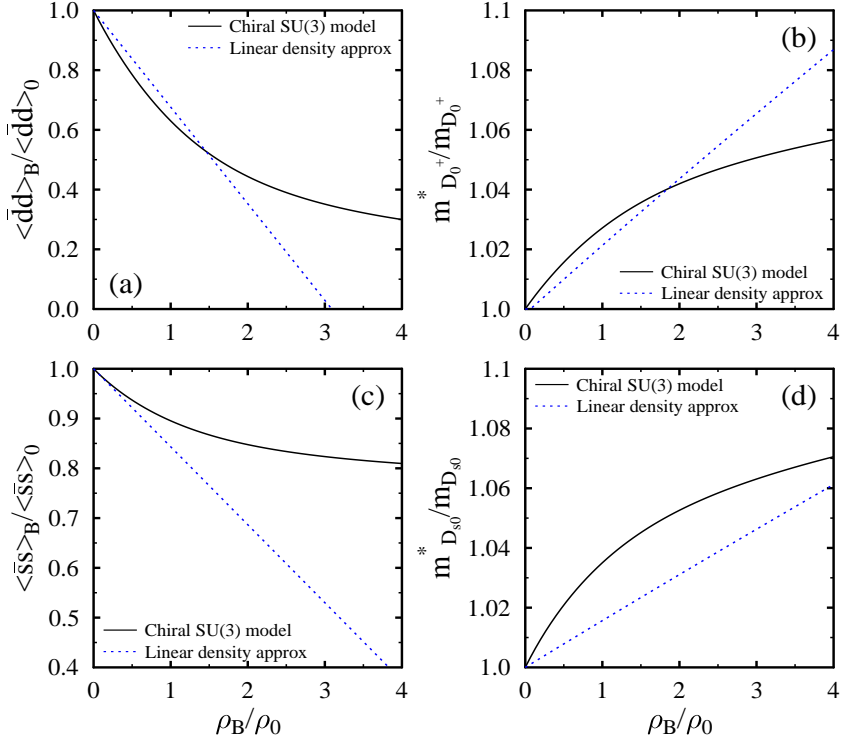}
\caption{Figure shows the variation of  in-medium mass of scalar $D_{0}$, $D_{s0}$ mesons and corresponding light quark, strange quark condensates (calculated using linear density approximation and chiral SU(3) model) as a function of baryonic density of the medium.}\label{fig:valid}
\end{figure}

%This behaviour of the masses and the decay constants of $D_0^0$, $D_0^+$ and $D_{s0}$ mesons, can be understood on the basis of the density and strangeness fraction dependence of the light quark and gluon condensates occurring in the \cref{qcdsum}.
%As we know that $\sigma$ and $\zeta$ fields contains light quarks $(u$ and $d)$ and strange  quark $s$ respectively. Therefore the former is more sensitive to the isospin asymmetric property of the medium, whereas $\zeta$ field to the presence of the hyperons in the medium. Therefore the behavior of these scalar fields (discussed in detail in Ref. \cite{Rahul}) are reflected through the medium modification in the light quark and gluon condensates and this is further reflected through the medium modification in the masses and decay constants of $D_0$ and $D_{s0}$ mesons.
%Further, in the present investigation,   we observe isospin splitting between $D_0^0$ and $D_0^+$ mesons. This is because of the finite value of $\delta$ field and  different masses of $u$ and $d$ quarks chosen in the present work. This further cause different  modification of $\left\langle \bar{u}u\right\rangle$ and $\left\langle \bar{d}d\right\rangle$ condensates as a function of isospin asymmetric parameter.

The finite isospin asymmetry of the medium causes the splitting in the in-medium masses  of $D_0^0$ and $D_0^+$ mesons.  For example,  in cold nuclear medium, at baryon density $\rho_B$=$\rho_0$, if we change isospin asymmetry parameter from $I$=0 to 0.5, the values of  masses and decay constants  of $D_0^0$ ($D_0^+$) mesons decrease (increase) by 0.3$\%$ (0.15$\%$) and 0.25$\%$ (0.1$\%$), respectively. At higher baryonic density, $4\rho_0$, above percentage values shift to 0.5$\%$ (0.2$\%$) and 0.6$\%$ (0.09$\%$), respectively.  
%Further, as mentioned earlier  in chiral SU(3) model the coupled equations of motion are solved for $\sigma$, $\zeta$, $\chi$ and $\delta$ fields, 
%%Further $\sigma$, $\zeta$ and $\chi$ fields change with the finite value of $\delta$ field in an isospin asymmetric medium.
%therefore, the value of $\sigma$, $\zeta$ field will also change with the finite value of the $\delta$ field in an isospin asymmetric medium.  
The change in isospin asymmetry of the medium also effect the in-medium masses of scalar $D_{s0}$ mesons.
For example, at baryonic density $\rho_0$ on shifting from $I = 0$ to $0.5$, we observed 0.3$\%$ (0.1$\%$)  decrease in the value of the mass (decay constant) of  $D_{s0}$ mesons at $T=0$ and $f_s$=0.  These percentage values further enhance to 0.7$\%$ (0.47$\%$), at higher baryonic density  $4\rho_0$.

%
% For example,  at baryonic density 4$\rho_0$, on moving from symmetric ($I$=0) to isospin asymmetric situations ($I$=0.5), the percentage change in the masses of $D_0^0$, $D_0^+$ and $D_{s0}$ mesons are observed as 6.3$\%$, 5.4$\%$  and 6$\%$, respectively from its vacuum values at $f_s$=0, $T$=0.  We can compare these values observed in the symmetric situations (keeping the other parameters constant) as 6.8$\%$, 5.2$\%$  and 6.5$\%$,  respectively. Similarly, in an isospin asymmetric nuclear medium, the percentage shift (from vacuum values) in the magnitude of decay constants are observed to be 5.3$\%$, 4$\%$  and 4$\%$, respectively, at $\rho_B$=4$\rho_0$, $T=0$. These values can be compared with  4.8$\%$, 4$\%$  and 3.6$\%$,  respectively obtained in symmetric situations. Clearly, isospin asymmetric situations cause an increase in the magnitude of the  masses and decay constants of $D_0^+$ meson,  whereas it cause decrease for $D_0^0$  mesons. 

%Here we emphasis on the fact that, we use chiral SU(3)  model to investigate the in-medium modification of  light quark and gluon condensates observed same as done for pseudo-scalar $D$ meson in our earlier work \cite{rahul2}.
In \cite{rahul2}, we observed negative shift in the masses of
pseudo-scalar $D$ meson using chiral SU(3)  model and QCD sum rules.
 The opposite  shift in the mass of scalar $D_0$ and pseudo-scalar $D$ meson is due to the opposite sign with the term $\frac{m_{c}\langle\bar{q}q\rangle_N}{2}$ \cref{qcdsum}, present in the Borel transformed equation (also see eq. (19) of \cite{rahul2}). This causes negative and positive value of the unknown parameter $a$ \cite{wang2}, calculated for scalar $D^0_0$  and pseudo-scalar $D$ mesons, respectively. This further cause positive and negative values of the scattering length for  $D_0^0 N$ and $D^0 N$, scattering, respectively. In \cref{fig:sle} we show the variation of scattering length 
 corresponding to scattering of $D_0^0$ and $D^0$ mesons with nucleons as a function of baryonic density  for isospin asymmetric parameters $I$=0 and 0.5, in cold nuclear medium.

Moreover, to understand more about the extent of isospin and flavour symmetry breaking in the medium, in \cref{fig:ratiod}, we plot the ratio of in-medium decay constants of $\frac{f^*_{D_0}}{f^*_{D_+}}$(subplot (a)), $\frac{f^*_{D_{s0}}}{f^*_{D^+}}$(subplot (b)) and  $\frac{f^*_{D_{s0}}}{f^*_{D^0}}$(subplot (c)) as a function of baryonic density at $T$=0. As expected, the ratio $\frac{f^*_{D^0_{0}}}{f^*_{D^+}}$ is more sensitive to the isospin asymmetry of the medium as compared to strangeness fraction. Opposite is true for the in-medium ratios of  $\frac{f^*_{D^0_{s0}}}{f^*_{D^0}}$ and $\frac{f^*_{D^0_{s0}}}{f^*_{D^+}}$.  

To check the reliability of the results of present work at higher value of baryonic density, in \cref{fig:valid}, we compare the in-medium behavior of the light quark condensates, $\left\langle \bar{d}d \right\rangle_{\rho_B}$ ($\left\langle \bar{s}s \right\rangle_{\rho_B}$) and the  in-medium mass of $D_0^+$ ($D_{s0}$) meson(in symmetric nuclear medium) both calculated  using the linear density approximation and chiral SU(3) model. Within linear density approximation, the light quark condensate $\left\langle \bar{d}d \right\rangle_{\rho_B}$, is calculated using
$\left\langle d\bar{d} \right\rangle _B$ = $\left\langle d\bar{d} \right\rangle _0$ + $\frac{\sigma_N \rho_B}{m_u+m_d}$,whereas the %where  $\sigma_N$ = $45$ MeV \cite{wang2}. On the other hand,
the strange quark condensate is calculated using $\left\langle \bar{s}s \right\rangle_{\rho_B}$   = 0.8$\left\langle \bar{q}q\right\rangle_{0}$+  $y\frac{\sigma_N \rho_B}{m_u + m_d}$, for $\sigma_N$ = 45 MeV and $m_u$+$m_d$ = 11 MeV \cite{wang2,hilger2009}. Here the term $\left\langle \bar{q}q\right\rangle_{0}$, is the vacuum value of light quark condensate and is given as (-0.245 GeV)$^3$. Also, the value of $y$ was taken to be 0.5. In addition, we calculate the mass of $D_0^+$ ($D_{s0}$) meson by considering  only condensate $\left\langle \bar{d}d \right\rangle_{\rho_B}$ ($\left\langle \bar{s}s \right\rangle_{\rho_B}$) in QCD sum rule equations,  which we  calculate using linear density approximation at zero temperature and symmetric nuclear medium. 
The linear behaviour of  light quark  and strange condensates is reflected in the linear variation of  masses of $D_0^+$ and $D_{s0}$ mesons. However, if we calculate  $\left\langle \bar{d}d \right\rangle_{\rho_B}$ ($\left\langle \bar{s}s \right\rangle_{\rho_B}$) using chiral SU(3) model, then we observe non-linear decrease as a function of baryonic density of the medium. Similarly, corresponding in-medium mass of $D_0^+$ ($D_{s0}$) meson increase non-linearly as a function of baryonic density. The observed non-linear decrease of the light quark condensate $\left\langle \bar{d}d \right\rangle_{\rho_B}$ at higher baryonic density of the medium, calculated using the chiral SU(3) model is in accordance of the work of \cite{kaiser}. In this work, authors calculated the light quark condensates beyond the linear density approximation using chiral perturbation theory. Therefore, the use of chiral SU(3) model to calculate the light quark condensates enables us to investigate the in-medium mass and decay constants of $D_0$ and $D_{s0}$ meson at higher baryonic density of the medium using QCD sum rules.

Additionally, we notice that the inclusion of the next to leading order  term (NLO) to the scalar quark condensates $\left\langle \bar{q}q\right\rangle$ in QCD sum rules (\cref{qcdsum})   enhances the magnitude of the shift in the mass of above mentioned meson \cite{wang2}. Further, we notice a  major contribution of the scalar quark condensates $\left\langle \bar{q}q\right\rangle$ to the shift in the mass of scalar $D_0$ and $D_{s0}$ meson as compared to the all other condensates. To understand this, we tabulate the numerical values of shift in the mass of  $D_0^+$ and $D_0^0$ meson in \cref{table_mass_indiv_scalar1,table_mass_indiv_scalar2}, respectively.  We also notice that the condensate $\langle \bar{q} i D_0 q\rangle _N$ which we do not calculate from the chiral SU(3) model has insignificant contribution on the shift in masses of above studied charmed mesons.

The uncertainties in the results of the present calculations may arise because of the medium modification in coupling constant, $g_{{{D_0/D_{s0}}N\Lambda_{\pi}}}$ and $g_{{{D_0/D_{s0}}N,\Sigma_{\pi}}}$ and the continuum threshold parameter $s_0$. In the present work, we neglect their in-medium modification. However, in symmetric nuclear medium, if we allow to decrease the value of coupling constant (continuum threshold parameter) by  5$\%$, then the shift in mass  of $D_0^0$ meson decrease (increase) by 1.5$\%$(15$\%$) at baryonic density $\rho_0$, and temperature, $T=0$. Likewise, the magnitude of shift in decay constant decrease (decrease) by 0.5$\%$(10$\%$). This indicates that the errors caused by the shift in value of coupling constant (continuum threshold parameter) may have insignificant (significant) effect on the shift in masses and decay constants of $D_0$ and $D_{s0}$ mesons. 

 \begin{table}
\begin{tabular}{|l|l|l|l|l|l|l|l|l|l|l|}
\hline
& & \multicolumn{4}{c|}{I=0} & \multicolumn{4} {c|}{I=0.5}\\  \cline{3-10}

$D_0^+$& & \multicolumn{2}{c|}{T=0} &\multicolumn{2}{c|}{T=100}& \multicolumn{2}{c|}{T=0} & \multicolumn{2}{c|}{T=100}    \\\cline{3-10}

& & $\rho_0$ & $4\rho_0$  &$\rho_0$&4$\rho_0$&$\rho_0$&4$\rho_0$ & $\rho_0$ & 4$\rho_0$ \\ \hline
All Condensates  &NLO&83&142&78 &140 &87 &145 & 82 & 141\\ \cline{2-10}
  & LO &64&125&58 &120 &68 &127 & 62 & 123 \\ \cline{1-10}
$\left\langle \bar{d}d \right\rangle _N$ $\neq$ 0  &NLO &84&145&78 &141 &89 &149 & 80 & 143\\  \cline{2-10}
  &LO & 63 &132 & 56&126 &66 &134 &60 &131\\  \hline
$\langle \bar{q} i D_0 q\rangle _N$ =0 &NLO&86&151&81 &146 &93 & 155 & 84 &146\\ \cline{2-10}
 & LO &66 &138 & 59 & 132 &70  &140  &63 &137\\ \hline

\end{tabular}
\caption{In the above table mass shift of $D_0^{+}$ mesons (in MeV) are compared by considering the contribution of individual condensates.}
\label{table_mass_indiv_scalar1}
\end{table}

\begin{table}
\begin{tabular}{|l|l|l|l|l|l|l|l|l|l|l|}
\hline
& & \multicolumn{4}{c|}{I=0} & \multicolumn{4} {c|}{I=0.5}\\  \cline{3-10}

$D_0^0$& & \multicolumn{2}{c|}{T=0} &\multicolumn{2}{c|}{T=100}& \multicolumn{2}{c|}{T=0} & \multicolumn{2}{c|}{T=100}    \\\cline{3-10}

& & $\rho_0$ & $4\rho_0$  &$\rho_0$&4$\rho_0$&$\rho_0$&4$\rho_0$ & $\rho_0$ & 4$\rho_0$ \\ \hline
All Condensates  &NLO&103&181&95 &170 &91 &166 & 88 & 164\\ \cline{2-10}
  & LO &87 &162 & 76& 156 & 78 & 148 & 72& 143 \\ \cline{1-10}
$\left\langle \bar{u}u \right\rangle _N$ $\neq$ 0  &NLO &105 &178 & 97 &168 &93 &164 &90 &161\\  \cline{2-10}
  &LO & 85 &173 & 75&165 &77 &156 &69 &153\\  \hline
$\langle \bar{q} i D_0 q\rangle _N$ =0 &NLO&104&184&92 &179 &99 & 171 & 89 &167\\ \cline{2-10}
 & LO &89 &180 & 79 & 173 &81  &164  &73 &160\\ \hline

\end{tabular}
\caption{In the above table mass shift of $D_0^{0}$ mesons (in MeV) are compared by considering the contribution of individual condensates.}
\label{table_mass_indiv_scalar2}
\end{table}
%\end{document}

Now, we compare the results of the present investigation with the available data of medium modification of scalar $D_0$ mesons. It should be noted that, no work is available in literature within any model which calculate the mass and decay constant of scalar $D_0$ and $D_{s0}$ mesons in strange hadronic matter. In Ref. \cite{wangs}, author applied linear density QCD sum rule and calculate the positive shift of 69 MeV for $D_0$ meson in cold symmetric nuclear matter.  In Ref. \cite{wang2} by adding the next to leading order term in the QCD sum rules, author found the shift in mass and decay constant of $D_0$ meson as 80 MeV and 11 MeV accordingly at cold and symmetric nuclear matter.   
 Furthermore, an extra widening of large width of the scalar $D_0$ mesons, whereas the width of nearly 100 MeV was observed for the case of $D_{s0}$ meson at normal nuclear matter by using coupled channel approach \cite{tolos}.  
    In Ref. \cite{hilger2009}, author observed the mass splitting between $D_0$ and $\bar{D_0}$ meson  by dividing the even and odd term  of correlation function in nuclear matter. However in the present work as mentioned earlier we observe average mass shift of $D_0$ and $\bar{D_0}$ meson by taking the average  particle and antiparticle current. 
 The results of the enhancement in the masses of scalar $D_0/D_{s0}$ meson  suggest us that scalar meson may not cause the $J/\psi$ suppression in the HIC experiments  and one might think that this enhanced mass of $D_0$ meson may act as facilitators to the production of $J/\psi$ state in heavy ion collision experiments. Also, the positive shift in mass may cause significant change in the values of absorption \cite{absorp} as well as production cross-section of higher charmonium states observed in HIC experiments \cite{cross}. Additionally, this in-medium enhancement may also be reflected through the measured values of the  elliptic flow, $v_2$, and nuclear modification factor, $R_{AA}$, of open charmed  mesons \cite{ahn}.  Further, the enhanced mass of scalar $D_0$ meson indicate the repulsive interactions of $D_0$ mesons with nucleons as well with the hyperons and therefore, the formation of scalar $D_0$ meson-nucleon/hyperons bound states may not be possible.

\begin{figure}
\centering
\includegraphics[width=14cm,height=12cm]{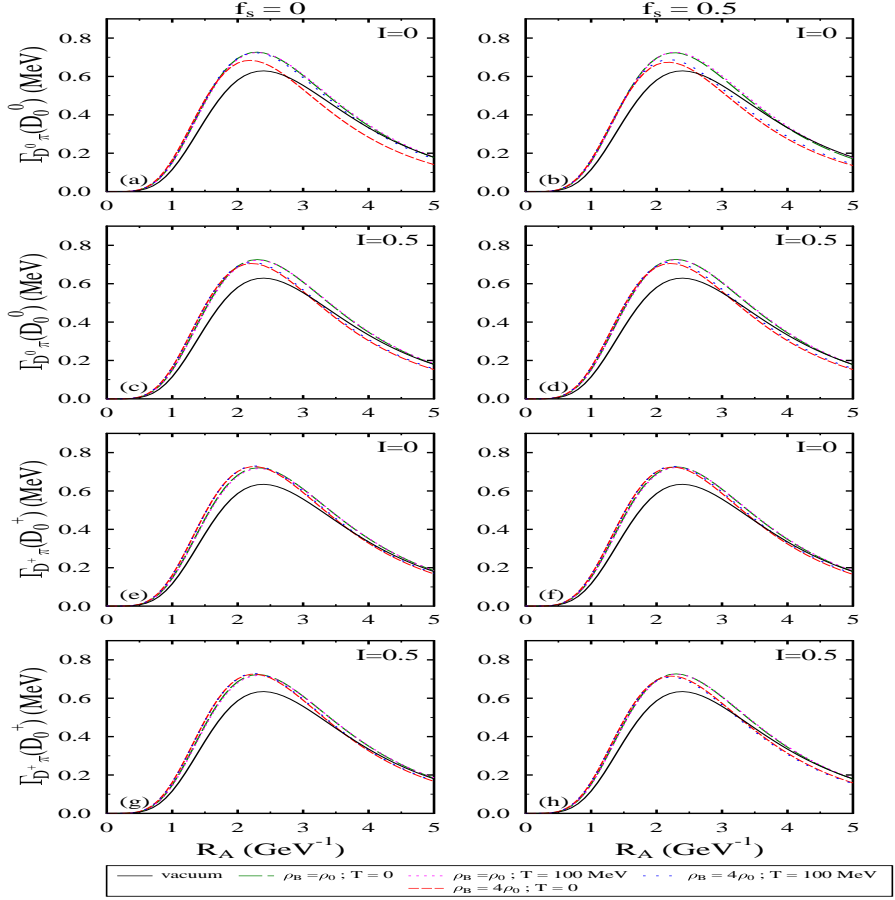}
\caption{Figure shows the variation of partial decay width of particular decay $D_0^+$  $\to$ $D^+$ + $\pi$ and $D_0^0$  $\to$ $D^0$ + $\pi$ as a function of  $R_A$ value (in GeV$^{-1}$).}\label{fig:D0}
\end{figure}
 
   \begin{figure}
\centering
\includegraphics[width=14cm,height=12cm]{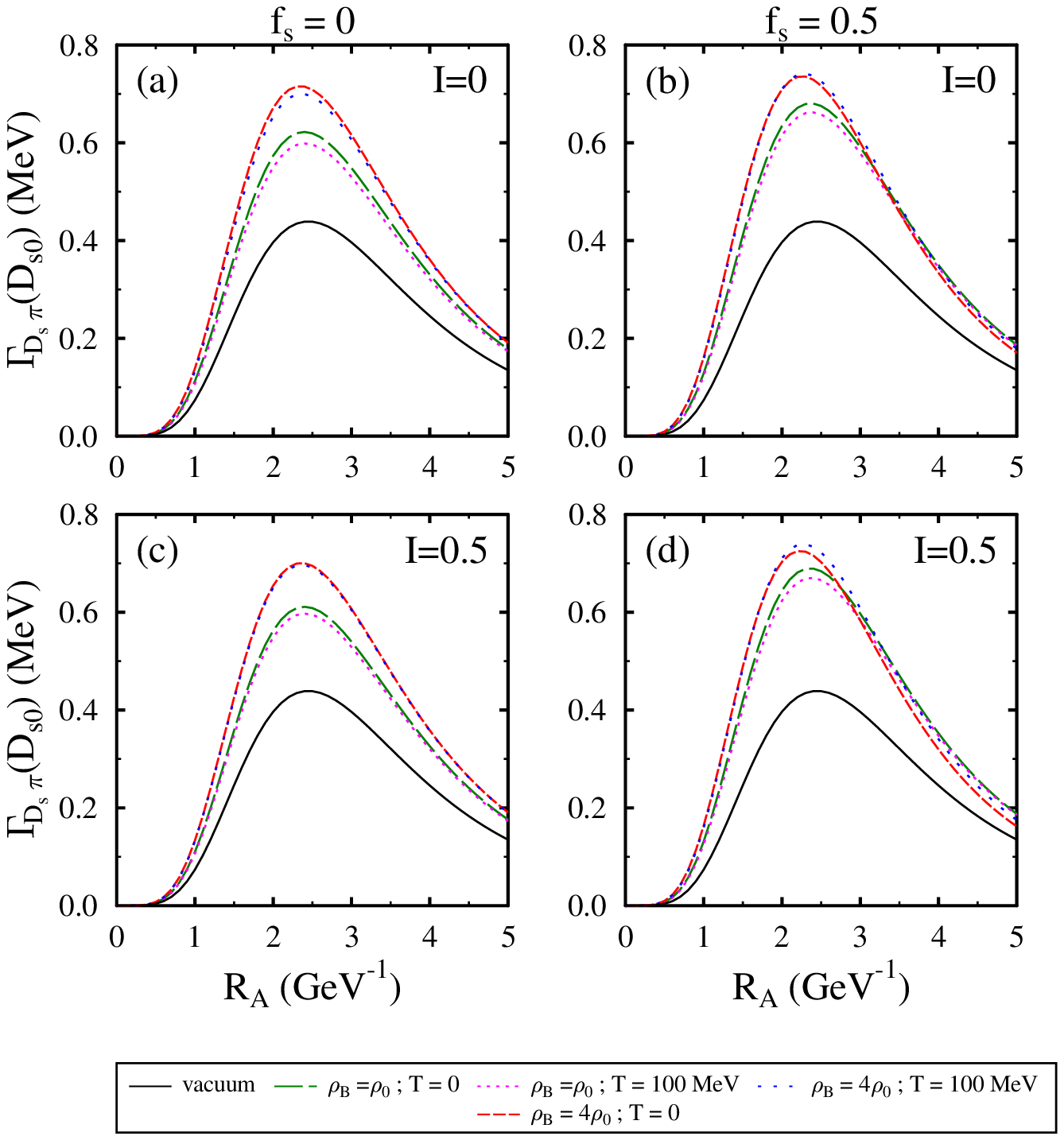}
\caption{Figure shows the variation of the partial decay width of a particular decay $D_{s0}$  $\to$ $D_s$ + $\pi$ as a function of respective $R_A$ value (in GeV$^{-1}$).}\label{fig:Ds0}
\end{figure}

 \subsection{In-medium partial decay width of $D_0^0$($D_0^+$) and $D_{s0}$ mesons}
 \label{sub_decay}
 
 In this section, using $^3P_0$ model, we shall calculate the in-medium partial decay width of the scalar $D^+_0$, $D^0_0$ and $D_{s0}$ mesons for the processes
$D_0^+$  $\to$ $D^+$ + $\pi$, $D_0^0$  $\rightarrow$ $D^0$ + $\pi$, and $D_{s0}$  $\rightarrow$ $D_s$ + $\pi$, respectively. In \cref{fig:D0}, we represent the partial decay widths $\Gamma_{D^+\pi}(D^+)$ and $\Gamma_{D^0\pi}(D^0)$, whereas in \cref{fig:Ds0}, we present $\Gamma_{D_{s}\pi}(D_{s0})$ as a function of $R_A$ values (where $A$ represents the parent meson). Here, as mentioned earlier, to calculate the in-medium partial decay width for the above mentioned processes, we consider the medium modified masses of parent as well as daughter mesons.
In above listed decay processes daughter mesons are pseudoscalar whereas parent are scalar. For the in-medium mass of pseudoscalar $D$ and $D_s$ mesons, we follow our earlier work \cite{rahul2}, where calculation were done using QCD sum rules and chiral SU(3) model.
Also, we include the medium modified mass of $\pi$ meson, calculated using chiral perturbation theory \cite{pionmass}. In \cite{pionmass}, authors studied the in-medium mass of $\pi$ mesons  in symmetric nuclear matter at zero temperature  including next to leading order term upto baryonic density 3$\rho_0$. 
As no work is still available on the study of mass shift
of $\pi$ mesons in asymmetric strange matter at finite temperatures,
therefore we use the same  shift in mass, for the isospin asymmetric strange hadronic matter also.

%Following this work, we used the values $-30$ and $-50$ MeV as the shift in mass of $\eta$ mesons at $\rho_0$ and $4\rho_0$, respectively.

%   Additionally, we include the medium modified mass of $\eta$ meson, calculated using heavy-baryon chiral perturbation theory combined with relativistic meanfield theory \cite{zhong1}. In \cite{zhong1} authors studied the in-medium mass of $\eta$ mesons  in zero temperature symmetric nuclear matter including next to leading order term upto baryonic density 3$\rho_0$. Therefore, we considered in-medium mass of $\eta$ meson calculated in \cite{zhong1}, along with the in-medium mass of $D_s$ mesons, and represent the partial decay widths of the processes $D_s^*(2715)$  $\to$ $D_s(1968)$ + $\eta$($D_s^*(2860)$  $\rightarrow$ $D_s(1968)$ + $\eta$)   in  \cref{fig:Ds(2715)}(\cref{fig:Ds(2860)}), as a function of their respective $R_{D_0}$ values.  
 
 The effect of in-medium modifications of parent and daughter mesons is observed to be significant on the partial decay width of $D_0$ and $D_{s0}$ mesons.  
%In the present work, we notice a significant influence of the  modification in the mass of respective mesons on their partial decay width, in addition to the respective $R_{A}$ values. 
From \cref{fig:D0} and \cref{fig:Ds0}, we notice an enhanced in-medium partial decay width for decays $D_0^+$  $\to$ $D^+$ + $\pi$, $D_0^0$ $\rightarrow$ $D^0$ + $\pi$, and $D_{s0}$  $\rightarrow$ $D_s$ + $\pi$ as compared to the vacuum values. Moreover, we do not observe any node in the above mentioned partial decay widths since the parent and daughter mesons are in their ground states.  Also, from \cref{eq:M,eq:G,eq:L} we note that the value of partial decay width is proportional to the square of decay amplitude, which   is further dependent on the spatial integral. Furthermore, this spatial integral  has been  solved analytically for the respective decay channel (\cref{I0,I1})  and therefore, behavior of the partial decay width is the resulting effect of the two integrals $I_0$ and $I_1$ occurring in \cref{eq:L}. Here through the competitive effect of the two integrals we observe the vacuum values of partial decay widths  $\Gamma_{D^+\pi}(D_0^+)$, $\Gamma_{D^0\pi}(D_0^0)$ and $\Gamma_{D_{s}\pi}(D_{s0})$ as 557 and 551 and 374 keV, respectively at $R_A$=1.89 GeV$^{-1}$ values. However, in symmetric nuclear medium, at $\rho_{B} = \rho_0$ and $T = 0$, the above values are observed to be 666, 653 and   544 keV, respectively. 
%Also on increasing the baryonic density to 4$\rho_0$ the values further increase to 650, .

Furthermore, on moving from symmetric nuclear ($f_s = 0$) to strange medium ($f_s = 0.5$), we observe enhancement in the respective values of  partial decay width and above listed values change to 669, 661 and 604 keV, respectively at $\rho_B$=$\rho_0$, and $T$ = 0. Here, we note that the in-medium mass of $D_{s0}$ meson is much sensitive to the finite strangeness fraction and therefore, the increase in the value of $\Gamma_{D_{s}\pi}(D_{s0})$ is more as compared to $\Gamma_{D^+\pi}(D_0^+)$ and $\Gamma_{D^0\pi}(D_0^0)$ in symmetric  strange hadronic matter.   
On the other hand, on increasing the temperature of the symmetric nuclear matter, the above mentioned partial decay width observed are observed as 660, 647 and 521 keV, respectively at normal nuclear matter density. 
 Furthermore, on moving from symmetric nuclear ($I = 0$) to asymmetric nuclear matter ($I = 0.5$) the above mentioned values shift to 662, 656 and 533 keV, respectively, for nuclear saturation density and zero temperature situation.  
Moreover, if we consider  $D_{s0}(2317)$ meson decaying to $D_s$ + $\pi$ mesons through $\eta$-$\pi^0$ mixing \cite{jlu}, then the observed vacuum values of the partial decay width was just 32 keV. Further, in normal nuclear matter density, $\rho_0$, and in cold symmetric nuclear medium, considering the mixing effect, the observed partial decay width enhance to 48 keV. This is because of enhanced decay channel caused by increase(decrease)in the masses of $D_{s0}$($D_s$) mesons.  Also, on addition of  hyperons along with the nucleons, at $\rho_B$=$\rho_0$, and $T = 0$, the decay width further increases to 56 keV. 

We shall now compare the results of the in-medium decay width with the previous works. As far as our knowledge regarding the literature is concerned, in-medium partial decay width of above mentioned process have not been evaluated so far. However, using the quark model authors has predicted the vacuum value of partial decay width of $P$-wave scalar $D_0$(2400) meson as 248 and 277 MeV, in ref. \cite{zhong2} and \cite{godfrey2}, respectively. Furthermore, in \cite{jlu}, authors used $^3P_0$ model to calculate the partial decay  width of $D_s(2317)$ meson through the $\eta$-$\pi^0$ mixing as 32 keV in vacuum.   Further, by taking $D_{s0}$ as four quark state authors observed its  partial decay width to $D_s \pi$ as  6 keV \cite{niel}. 
 Also, in \cite{bard} above mentioned width was observed as 21.5 keV,  using full chiral theory on equating the mass gap  of $0^+$ and $1^+$ states with $0^-$ and $1^-$ states.  Moreover, by considering $D_{s0}$ state as $s^p_l$ =1/2$^+$ and using heavy quark symmetries along with Vector Meson Dominance ansatz, authors observed the value of $\Gamma_{D_s \pi} (D_{s0})$ $\simeq$ 7 keV \cite{col}. 
 Similar results of partial decay width of above mentioned process were  observed as 10, 16 and 39 keV in Ref. \cite{godfr}, \cite{faya} and \cite{wei}, respectively.
%  Clearly, these results are in agreement with the present results of partial decay width of   $\Gamma_{D_s \pi} (D_{s0})$. 
From the above discussion we observe that the partial decay widths of scalar $D_0$ and $D_{s0}$ mesons are quite model dependent and need to verify in the future experiments.
%Here we emphasis on the fact that, the accurate measurement of the partial decay widths and their medium behavior will be verified in the future HIC at FAIR facilities, and this  will further validate the theoretical model and the proper state of the excited $D$ and $D_s$ mesons. 

 \section{Summary}
 \label{summary} We observed the positive(negative) shift in masses(decay constants) of scalar $D_0(2400)$ and $D_{s0}(2317)$ mesons, using chiral SU(3) model and QCD sum rules. Positive shift in mass of these mesons indicate that these meson may act as facilitators to the production of $J/\psi$ state in HIC experiments.
This is because the threshold value of $D\bar{D}$ pair will be above the mass of excited charmonium states and thus, these charmonium states will decay to $J/\psi$  mesons instead of $D\bar{D}$ pairs. 
  Also, these repulsive interactions indicate that the scalar $D_0/D_{s0}$ meson may not form bound states with the nucleons as well as with hyperons.
   The enhanced value of in-medium mass may have significant effect on the elliptic flow and nuclear modification factor, $R_{AA}$ of these scalar open charmed meson produced in HICs experiments.
   Furthermore, we take the in-medium mass of these scalar $D$ mesons as an application in $^3P_0$ model and evaluate their in-medium partial decay widths for the processes, $D_0(2400)$ $\to$ $D+\pi$ and $D_{s0}^*(2317)$ $\to$ $D_s+ \pi$. We observe that, as the mass of scalar $D$ meson increase in the hyperonic (along with the nucleons) medium, this results in the significant increase in the corresponding partial decay widths. 
%In the present paper, we also take the in-medium mass of  pseudoscalar $D$ and $\pi$ mesons.  

% However, even at baryonic density 4$\rho_0$, the observed partial decay width never appear to cross the vacuum $DK$ channel.  Therefore, to calculate the exact quantum numbers of $D_s^*(2715)$ and $D_s^*(2860)$ states, exact knowledge of their in-medium masses and partial decay widths of all the possible decay channel, as well as more accurate data from the experiments is needed.  
 
\acknowledgments
 The authors gratefully acknowledge the financial support from
the Department of Science and Technology (DST), Government of India for research project under
 fast track scheme for young scientists (SR/FTP/PS-209/2012).

\end{document}